\documentclass{bmcart}
\usepackage{amsmath}
\usepackage[super,sort&compress,comma]{natbib}
\usepackage{hyperref}
\hypersetup{ colorlinks,
    citecolor=blue,
    filecolor=blue,
    linkcolor=blue,
    urlcolor=blue
}
\usepackage{graphicx}

\usepackage{subcaption}
\usepackage{array}
\newcolumntype{L}[1]{>{\raggedright\let\newline\\\arraybackslash\hspace{0pt}}m{#1}}
\newcolumntype{C}[1]{>{\centering\let\newline\\\arraybackslash\hspace{0pt}}m{#1}}
\newcolumntype{R}[1]{>{\raggedleft\let\newline\\\arraybackslash\hspace{0pt}}m{#1}}

\usepackage[bmc]{bmcart-biblio}

\usepackage[polish,english]{babel}
\usepackage[utf8]{inputenc}
\usepackage[T1]{fontenc}
\usepackage{caption}
\usepackage{lineno}
\newcommand{\Left}{\textbf{Left:} }
\newcommand{\Right}{\textbf{Right:} }
\newcommand{\Middle}{\textbf{Middle:} }

\begin{document}

%%% Start of article front matter
\begin{frontmatter}

\begin{fmbox}
\dochead{Research}
\title{Performance assessment of the 2$\gamma$ positronium imaging with the
  total-body PET scanners }

%%%%%%%%%%%%%%%%%%%%%%%%%%%%%%%%%%%%%%%%%%%%%%
%%                                          %%
%% Enter the authors here                   %%
%%                                          %%
%% Specify information, if available,       %%
%% in the form:                             %%
%%   <key>={<id1>,<id2>}                    %%
%%   <key>=                                 %%
%% Comment or delete the keys which are     %%
%% not used. Repeat \author command as much %%
%% as required.                             %%
%%                                          %%
%%%%%%%%%%%%%%%%%%%%%%%%%%%%%%%%%%%%%%%%%%%%%%

%\author[
%   addressref={aff1},                   % id's of addresses, e.g. {aff1,aff2}
%   corref={aff1},                       % id of corresponding address, if any
%   noteref={n1},                        % id's of article notes, if any
%   email={jane.e.doe@cambridge.co.uk}   % email address
%]{\inits{JE}\fnm{Jane E} \snm{Doe}}
\author[
   addressref={UJ},
]{\fnm{P.}~\snm{Moskal}}
\author[
   addressref={UJ},
   email={dk.dariakisielewska@gmail.com}
]{\fnm{D.}~\snm{Kisielewska}}
\author[
   addressref={UJ},
]{\fnm{Z.}~\snm{Bura}}
\author[
   addressref={UJ},
]{\fnm{C.}~\snm{Chhokar}}
\author[
   addressref={INFN},
]{\fnm{C.}~\snm{Curceanu}}
\author[
   addressref={UJ},
 ]{\fnm{E.}~\snm{Czerwi{\'n}ski}}
\author[
   addressref={UJ},
]{\fnm{M.}~\snm{Dadgar}}
\author[
   addressref={UJ},
]{\fnm{K.}~\snm{Dulski}}
\author[
   addressref={PAN},
]{\fnm{J.}~\snm{Gajewski}}
\author[
   addressref={UJ},
]{\fnm{A.}~\snm{Gajos}}
\author[
   addressref={UMCS},
]{\fnm{M.}~\snm{Gorgol}}
\author[
   addressref={INFN},
]{\fnm{R.}~\snm{Del~Grande}}
\author[
   addressref={UV},
]{\fnm{B.~C.}~\snm{Hiesmayr}}
\author[
   addressref={UMCS},
 ]{\fnm{B.}~\snm{Jasi{\'n}ska}}
\author[
   addressref={UJ},
]{\fnm{K.}~\snm{Kacprzak}}
\author[
   addressref={UJ},
 ]{\fnm{A.}~\snm{Kami{\'n}ska}}
\author[
   addressref={UJ},
 ]{\fnm{{\L}.}~\snm{Kap{\l}on}}
\author[
   addressref={UJ},
]{\fnm{H.}~\snm{Karimi}}
\author[
   addressref={UJ},
]{\fnm{G.}~\snm{Korcyl}}
\author[
   addressref={NCBJ},
]{\fnm{P.}~\snm{Kowalski}}
\author[
   addressref={UJ},
]{\fnm{N.}~\snm{Krawczyk}}
\author[
   addressref={NCBJ-HE},
 ]{\fnm{W.}~\snm{Krzemie{\'n}}}
\author[
   addressref={UJ},
]{\fnm{T.}~\snm{Kozik}}
\author[
   addressref={UJ},
]{\fnm{E.}~\snm{Kubicz}}
\author[
   addressref={UJCM},
 ]{\fnm{P.}~\snm{Ma{\l}czak}}
\author[
   addressref={UJ,Mosul},
]{\fnm{M.}~\snm{Mohammed}}
\author[
   addressref={UJ},
 ]{\fnm{Sz.}~\snm{Nied{\'z}wiecki}}
\author[
   addressref={UJ},
 ]{\fnm{M.}~\snm{Pa{\l}ka}}
\author[
   addressref={UJ},
 ]{\fnm{M.}~\snm{Pawlik-Nied{\'z}wiecka}}
\author[
   addressref={UJCM},
 ]{\fnm{M.}~\snm{P\k{e}dziwiatr}}
\author[
   addressref={NCBJ},
 ]{\fnm{L.}~\snm{Raczy{\'n}ski}}
\author[
   addressref={UJ},
]{\fnm{J.}~\snm{Raj}}
\author[
   addressref={PAN},
 ]{\fnm{A.}~\snm{Ruci{\'n}ski}}
\author[
   addressref={UJ},
]{\fnm{S.}~\snm{Sharma}}
\author[
   addressref={UJ},
]{\snm{Shivani}}
\author[
   addressref={NCBJ},
]{\fnm{R.~Y.}~\snm{Shopa}}
\author[
   addressref={UJ},
]{\fnm{M.}~\snm{Silarski}}
\author[
   addressref={UJ,INFN},
]{\fnm{M.}~\snm{Skurzok}}
\author[
   addressref={UJ},
 ]{\fnm{E.~{\L}.}~\snm{St{\k{e}}pie\'n}}
\author[
   addressref={Ghent},
]{\fnm{S.}~\snm{Vandenberghe}}
\author[
   addressref={NCBJ-HE},
 ]{\fnm{W.}~\snm{Wi{\'s}licki}}
\author[
   addressref={UMCS},
 ]{\fnm{B.}~\snm{Zgardzi{\'n}ska}}

%%%%%%%%%%%%%%%%%%%%%%%%%%%%%%%%%%%%%%%%%%%%%%
%%                                          %%
%% Enter the authors' addresses here        %%
%%                                          %%
%% Repeat \address commands as much as      %%
%% required.                                %%
%%                                          %%
%%%%%%%%%%%%%%%%%%%%%%%%%%%%%%%%%%%%%%%%%%%%%%

\address[id=UJ]{%                           % unique id
  \orgname{Faculty of Physics, Astronomy and Applied Computer Science, Jagiellonian University}, % university, etc
  \street{prof. Stanis{\l}awa {\L}ojasiewicza 11},                     %
  \postcode{30-348}                                % post or zip code
  \city{Cracow},                              % city
  \cny{Poland}                                    % country
}
\address[id=INFN]{%
  \orgname{INFN, Laboratori Nazionali di Frascati},
  %\street{},
  \postcode{00044}
  \city{Frascati},
  \cny{Italy}
}
\address[id=PAN]{%
  \orgname{Institute of Nuclear Physics PAN},
  %\street{},
  %\postcode{}
  \city{Cracow},
  \cny{Poland}
}
\address[id=UMCS]{%
  \orgname{Institute of Physics, Maria Curie-Sk\l odowska University},
  %\street{},
  \postcode{20-031}
  \city{Lublin},
  \cny{Poland}
}
\address[id=UV]{%
  \orgname{Faculty of Physics, University of Vienna},
  %\street{},
  \postcode{1090}
  \city{Vienna},
  \cny{Austria}
}

\address[id=NCBJ]{%
  \orgname{Department of Complex Systems, National Centre for Nuclear Research},
  %\street{},
  \postcode{05-400}
  \city{Otwock-\'Swierk},
  \cny{Poland}
}
\address[id=NCBJ-HE]{%
  \orgname{High Energy Physics Division, National Centre for Nuclear Research},
  %\street{},
  \postcode{05-400}
  \city{Otwock-\'Swierk},
  \cny{Poland}
}
\address[id=UJCM]{%
  \orgname{2nd Department of General Surgery, 
  Jagiellonian University Medical College},
  %\street{Kopernika 21},
  %\postcode{31-501}
  \city{Cracow},
  \cny{Poland}
}
\address[id=Mosul]{%
  \orgname{Department of Physics, College of Education for Pure Sciences, University of Mosul},
  %\street{},
  %\postcode{}
  \city{Mosul},
  \cny{Iraq}
}
\address[id=Ghent]{%
  \orgname{Department of Electronics and Information Systems, MEDISIP, Ghent University-IBiTech},
  \street{De Pintelaan 185 block B},
  \postcode{B-9000}
  \city{Ghent},
  \cny{Belgium}
}

\begin{artnotes}
%\note{Sample of title note}     % note to the article
%\note[id=n1]{Equal contributor} % note, connected to author
\end{artnotes}

\end{fmbox}% comment this for two column layout

\begin{abstractbox}
\begin{abstract}
  \parttitle{Purpose} 
  In living organisms the positron-electron annihilation (occurring during the PET imaging) proceeds in about 30\% via creation of a  metastable ortho-positronium atom.  In the tissue,  due to the pick-off and conversion processes, over 98\% of ortho-positronia annihilate into two 511~keV photons. In this article we assess the feasibility for reconstruction of the mean ortho-positronium lifetime image based on annihilations into  two photons. The main objectives of this work include:  (i)  estimation of the sensitivity of the total-body PET scanners  for the ortho-positronium mean lifetime imaging using $2\gamma$ annihilations, and (ii) estimation of the spatial and time resolution of the ortho-positronium image as a function of the coincidence resolving time (CRT) of the scanner. \\

  \parttitle{Methods} 
  Simulations are conducted assuming that radiopharmaceutical is labelled with $^{44}Sc$ isotope
  emitting one positron and  one prompt gamma.  The image is reconstructed on the basis of triple
  coincidence events.
   The ortho-positronium lifetime spectrum is determined for each voxel of the image.   Calculations were performed for cases of  total-body detectors build of (i)  LYSO scintillators as used in the EXPLORER PET, and  (ii) plastic scintillators as anticipated for the cost-effective total-body J-PET scanner.  To assess the spatial and time resolution the three cases were considered assuming that CRT is equal to 140~ps, 50~ps and 10~ps. 

  \parttitle{Results}  The estimated total-body PET sensitivity  for the registration and selection
  of image forming triple coincidences ($2\gamma + \gamma_{prompt}$) is larger by a factor of 12.2 (for LYSO PET) and by factor of 4.7 (for plastic PET) with respect to the  sensitivity for the standard $2\gamma$ imaging by LYSO PET scanners with AFOV~=~20~cm. The spatial resolution of the ortho-positronium image is comparable with the resolution achievable when using TOF-FBP algorithms already for CRT~=~50~ps. For the 20~min scan the resolution better than 20~ps is expected for the  mean ortho-positronium lifetime image determination.

  \parttitle{Conclusions} 
  Ortho-positronium mean lifetime imaging based on the annihilations into two photons and prompt
  gamma is shown to be feasible with the advent of the high sensitivity total-body PET systems and
  time resolution of the order of tens of picoseconds. \\
\end{abstract}


\end{abstractbox}

\end{frontmatter}

\section{Background}
\label{sec::introduction}
\begin{figure}[h]
  \centering
    \includegraphics[width=0.6\textwidth]{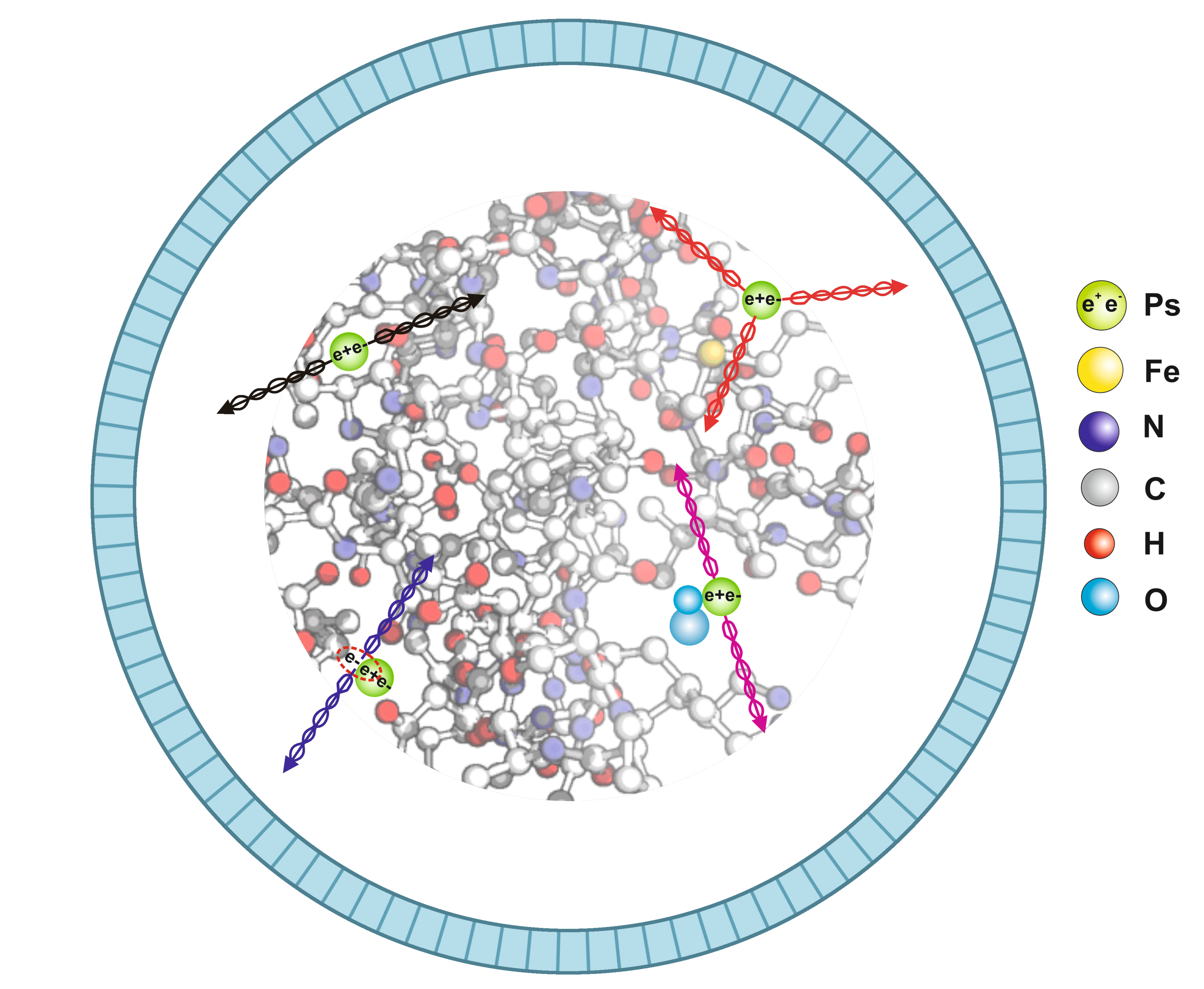}
    \caption{
      Pictorial representation of the single detection ring of the positron emission tomography
      scanner and (not to scale) magnified part of the hemoglobin molecule with pictorial
      representation of the possible ways of decays of positronium atoms (Ps) trapped in the
      intramolecular voids. Left-upper (black arrows) and right-upper (red arrows) indicate
      annihilations in the space free of electrons for para-positronium and ortho-positronium,
      respectively. 
      Annihilation of positronium through the interaction with the
      electron from the surrounding molecule is shown in the left-lower corner (violet arrows) while in the right-lower part the conversion of ortho-positronium into
      para-positronium via interaction with the oxygen molecule and subsequent decay of para-positronium
      to two photons (magenta arrows) are presented~\cite{Moskal2019NatureReview}.  
     \label{fig::detector}
   }
\end{figure}

In the state-of-the art positron emission tomography (PET) the density distribution of points of
positron-electron annihilations is used for the determination of the image of standardised uptake
value (SUV) of the radiopharmaceutical administered to the patient. The current  PET diagnostics
does not take advantage of the fact that positron and electron may form a positronium atom. Yet,
in up to about 40\% cases~\cite{Moskal_2019_PMB,Harpen2004}, positron-electron annihilations inside
the human body proceeds via the creation of the metastable positronium atom which in turn in quarter of
cases appears as para-positronium (pPs) decaying to two photons and in three quarter of cases as
ortho-positronium (oPs) decaying in vacuum into three photons. When trapped in the body, the
ortho-positronium   creation probability and  mean lifetime strongly depend on the tissue's
nanostructure and the concentration of bio-active molecules (e.g. free radicals, reactive oxygen
species, and antioxidants) which can interact with the emitted positrons as well as with the formed
positronium~\cite{Moskal_2019_PMB}.  
In particular ortho-positronium lifetime depends significantly on the size of free volume between
atoms whereas its formation probability depends on their concentration. While both lifetime and
formation probability depend on the concentration and type of biofluids and bio-active
molecules~\cite{Moskal2019NatureReview}. Therefore, these ortho-positronium properties may be
considered as diagnostic indicators complementary to the presently available SUV
index~\cite{Moskal_2019_PMB,patentMoskal}. Recently the in-vitro studies indicated that indeed
positronium mean lifetime and its production probability as well as the average time of direct
annihilation differ for healthy and cancerous uterine tissues operated from the
patients~\cite{JasinskaActaB:2017,JasinskaActaA:2017}. Another in-vitro measurements, performed with
blood taken from patients before and after the chemotherapy or radiotherapy, demonstrated
dependence
of positronium properties in blood on the time after anti-neoplastic therapy~\cite{Pietrzak}.
In Figure~\ref{fig::detector}, an example of the hemoglobin molecule
and the main mechanisms for the
positronium annihilations inside the cells is presented.
The positron annihilation lifetime spectroscopy (PALS) is a well established method in the material
science~\cite{Hugenschmidt2014,Reisfeld2016,Chen2007}.
However, in order to make use of the positronium properties in the in-vivo medical diagnostics 
development of the system combining PET and PALS is
required~\cite{Moskal_2019_PMB,Moskal2019NatureReview}.

Recently, a method of positronium mean lifetime imaging, in which the lifetime, production
probability and position of positronium  is determined on an event-by-event basis using
$oPs\to3\gamma$ decay was described~\cite{Moskal_2019_PMB,patentMoskal}. The place and time of
ortho-positronium decay is reconstructed by application of the trilateration
technique~\cite{Gajos:2016nfg,phdAlek} which uses times and positions of registrations of three photons from
$oPs\to3\gamma$ decay. The method is applicable for radiopharmaceuticals labeled with  $\beta^+$
decaying isotope emitting prompt gamma (e.g. $^{44}$Sc)~\cite{Szkliniarz2015,SITARZ2020108898}, where prompt gamma is used to determine the time of the ortho-positronium formation.  
The $3\gamma$ events were chosen because in case of the registration of the 3$\gamma$ in principle one can reconstruct the annihilation point with better spatial precision with respect to the case when only two photons are available. It was shown~\citep{Moskal_2019_PMB} that for the total-body J-PET scanner, when administering to the patient a typical activity of 370~MBq, one can expect about 700 registered events per cubic centimeter of the examined patient after 20 minutes of data collection. This is a promising result however the expected statistics is rather low.
Low rate of $oPs\to 3\gamma$ decays inside the tissue is due to the interaction of positronium with
electrons from the surrounding atoms (pick-off process~\citep{Garvin1571}) and due to the
conversion~\cite{Consolati1998,Ferrell1958,Zgardzinska2015} of ortho-positronium into para-positronium via interactions with the bio-active molecules (Fig.~\ref{fig::detector}). Because of these processes the ortho-positronium lifetime decreases in the tissue to the range of few nanoseconds~\cite{JasinskaActaB:2017,JasinskaActaA:2017} and the fraction of its decay rate into three photons $f_{oPs\to3\gamma}$ decreases from $f_{oPs\to3\gamma}=1$ in vacuum to 
$f_{oPs\to3\gamma}$= $\tau_{tissue}/\tau_{vacuum}$  in the tissue~\cite{Jasinska-Moskal2017}, where
$\tau_{tissue}$ and $\tau_{vacuum} = 142$~ns denote the ortho-positronium mean lifetime in the tissue and in vacuum, respectively. E.g. for $\tau_{tissue}~=~2~ns$, the ortho-positronium decays 70 times more frequent to 2$\gamma$ than to 3$\gamma$.

Lifetime of the decaying object may be determined by the measurement of any of its decay channels and hence  the ortho-positronium mean lifetime imaging can be performed based on the $oPs\to 3\gamma$ decay~\cite{Moskal_2019_PMB}, as well as based on the pick-off and conversion processes leading to the two back-to-back photons. 
In this article, we assess the feasibility of the $2\gamma$ ortho-positronium lifetime imaging for
the total-body PET scanners assuming that the  radiopharmaceutical is labeled with $^{44}Sc$ isotope
emitting positrons and prompt photon with energy of 1160~keV and using two back-to-back photons for
the reconstruction of the ortho-positronium decay time and decay position. The prompt photon is used
to determine the time of the creation of positronium. Reconstruction of the time difference between
annihilation and emission of the positron enables to disentangle between processes when
para-positronium decays to two photons (black arrows in Fig.~\ref{fig::detector}) and
ortho-positronium converts to two photons~(magneta and violet arrows in Fig.~\ref{fig::detector}).
The lifetime of para-positronium (equal to 125~ps in vacuum) does not alter much as a function of
properties of the tissues nanostructure (reaches about 230~ps) whereas the lifetime of ortho-positronium varies in the tissue in the range of few nanoseconds~\cite{JasinskaActaB:2017} and may occur to be useful as a diagnostic indicator~\cite{Moskal_2019_PMB,Moskal2019NatureReview}.

A statistical method of lifetime image reconstruction are yet to be conceived. Therefore, at present, in the event-by-event technique of ortho-positronium mean lifetime imaging the spatial precision of a single annihilation event reconstruction is equivalent to the spatial precision of the image. Hence, the resolution of the ortho-positronium mean lifetime image will directly depend on the time resolution of the PET detector. 
At present the newest TOF-PET scanners are characterized by the TOF resolution of about
210~ps~\cite{Vission} corresponding to the spatial resolution along the line of response (LOR) of
about 3.8~cm. Recently a detector design with  SiPM has been reported, with CRT = 85~ps for $2
\times 2 \times 3 \mbox{ mm}^3 $ LSO:Ce doped with 0.4\%Ca crystals and CRT of  140~ps for $2  \times 2  \times 20 \mbox{ mm}^3$ crystals with the length as used in the
current PET devices~\citep{PMB_Nemallapudi}.
Thus the TOF resolution, and consequently  spatial resolution for a single event, is gradually improving 
by development of new crystals, SiPMs~\cite{Bisogni2019}, fast high frequency
electronics~\cite{Gundacker2019}, signals filtering~\cite{Bieniosek_2016}
applications of the time ordered statistics~\cite{Seifert2012,Moskal:2016ztv}, signal waveform
sampling~\cite{Berg2017,Kyu2018} including fast and cost-effective sampling in voltage
domain~\cite{Palka:2017wms}, and advent of machine learning techniques~\cite{Zhu2018}.

Most recently, for the single detectors 30~ps resolution was achieved  which is equivalent to
position resolution of 4.5~mm along the line of response~\cite{Ota_2019}, and there is a continuous
effort to improve it further even down to 10~ps~\cite{Lecoq:2017,Gundacker2019} which would enable to reconstruct the $2\gamma$ annihilation point with precision of
1.5~mm, where the position and time information will be directly available for each decay event. In
this context it is worth mentioning that recently a  new quenching circuit (QC) and single photon
avalanche diode (SPAD) technology was introduced with 7.8~ps resolution~\cite{Nolet2018}
resulting in the resolution of 17.5~ps for the full chain of SiPM with QC and
TDC~\cite{NOLET201829}. 

In this article, based on the Monte-Carlo simulations, we argue that with the time resolution in the
order of tens of picoseconds and the advent of the high sensitivity total-body PET
systems~\cite{Badawi2019,Surti_2018} the $2\gamma~+~\gamma_{prompt}$ mean lifetime positronium imaging based on time measurements is
becoming realistic in the nearest future.  

In the next section the main assumptions applied in the simulations are presented. Further on, the sensitivity for the simultaneous registration of the back-to-back photons from positronium decay and prompt photon including selection of the image forming events is estimated for the total-body PET scanners built from LYSO crystal as well as for the cost effective version of the total-body PET built from plastic scintillators. 
Next, results of detailed Monte-Carlo simulations of the response of the J-PET total body scanner to the point like sources arranged in the configuration as described in the NEMA norm are performed 
and the regular PET 2$\gamma$ annihilation images as well as ortho-positronium mean lifetime images are reconstructed and 
compared for the three cases of assumed coincidence resolving time (CRT) of 140~ps, 50~ps and 10~ps.  
Finally it is shown that, owing to the large axial 
field-of-view of the total body scanners, the sensitivity for the positronium lifetime imaging is
even larger than the present sensitivity for the $2\gamma$ metabolic imaging with the PET scanners having 20~cm axial length, even though the discussed positronium lifetime imaging requires registration of triple coincidence events including the prompt gamma and the two back-to-back photons. %originating from the ortho-positronium pick-off and conversion processes. 

\section{Materials and Methods}
\label{sec::materials}

\begin{figure}[tp]
  \centering
    \includegraphics[width=0.9\textwidth]{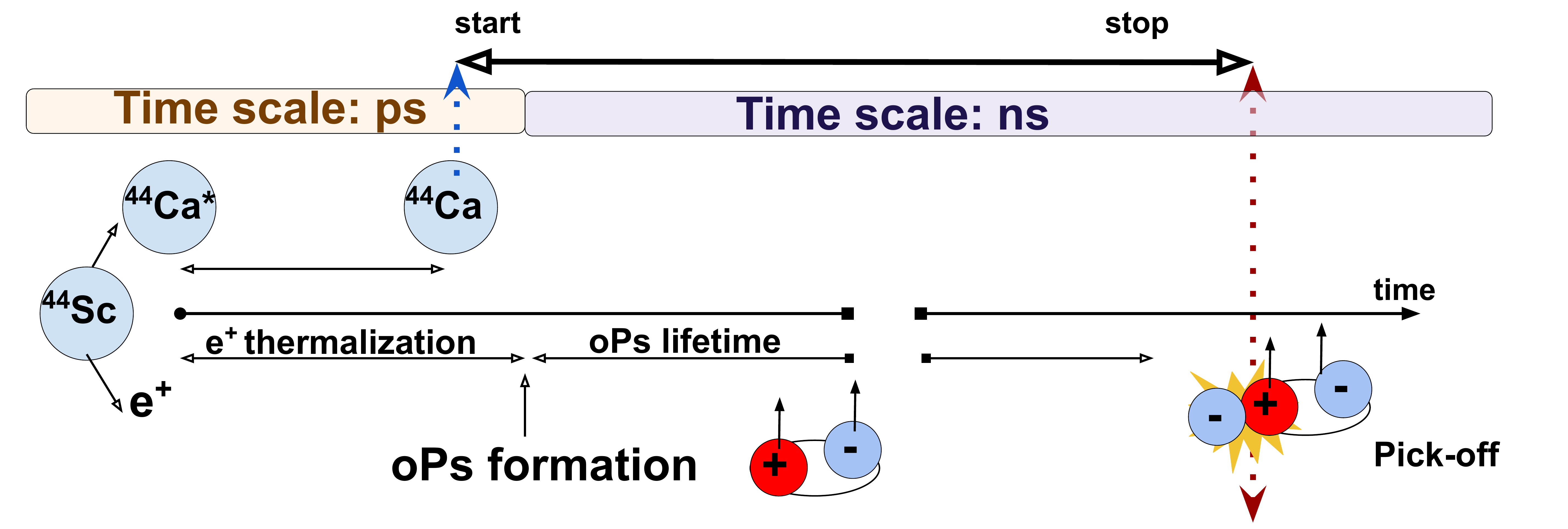}
    \caption{Scheme of the time sequence in the processes used for positronium imaging. A $^{44}$Sc
    nucleus undergoes $\beta^+$ decay. Next, on the average after about 3~ps, excited $^{44}$Ca$^*$
    emits prompt gamma with energy of 1160~keV (dotted blue arrow). 
    Parallelly, positron travels through matter, thermalizes and forms an ortho-posironium bound
    state. Interaction with surrounding molecules or conversion process leads to emission of two
    photons. The mean ortho-positronium lifetime is in the order of nanoseconds, in contrast to the
    duration of thermalization~\cite{Champion2005} and deexcitation~\cite{NUCLEIDE:LARA} processes which are in the order of ten picoseconds.
        \label{fig::pickofftimescale}
   }
\end{figure}

Positronium mean lifetime imaging may be defined as spatially resolved determination of the ortho-positronium lifetime 
inside the patient's body. For this purpose it is required to determine the lifetime and position of ortho-positronium 
on the event-by-event basis thus enabling calculation of the mean lifetime of ortho-positronium for each voxel of the image.
The simulations presented in this article are conducted assuming that radiopharmaceutical is labeled with $^{44}Sc$ isotope which emits positron and a remaining excited daughter nucleus $^{44}Ca^*$ emits prompt gamma with energy of 1160~keV  via reaction chain:
$^{44}Sc \to ^{44}\!\!Ca^* \ e^+ \ \nu \to ^{44}\!\!Ca \ \gamma \ e^+ \ \nu$  (see
Fig.~\ref{fig::pickofftimescale}). The single (triple coincident) event used for imaging contains three signals that
carry information about the position and time of the photons' interaction in the detector: with one signal corresponding to prompt gamma and two signals
corresponding to 511~keV photons originating from the annihilation of ortho-positronium while
interacting with the molecular environment.  The times and positions of interactions of 511~keV
photons are used to determine the time and position of the annihilation point, whereas the time and
position of the prompt gamma enable to determine the moment of the deexcitation of the $^{44}$Ca$^*$
nucleus which with the precision of tens of picoseconds~\cite{nncd}
can be associated with the moment of the creation of positronium.  The lifetime spectrum determined
for each voxel of the patient enables to extract information about the mean ortho-positronium
lifetime and its production
probability~\cite{patentMoskal,Dulski2017}. 

\subsection{Analytic estimation of sensitivity}
\label{sec::analytic}

First we performed simplified simulations in order to estimate a sensitivity for the registration
and selection of triple coincidences useful for the positronium image reconstruction. In the
simulation it was assumed that the PET scanner is built from the cylindrical layer of scintillator
detectors. Calculations were performed for the case of the LYSO scintillators with thickness of
d~=~1.81~cm as used in the EXPLORER~\cite{Leung_2018} and plastic
scintillators with thickness of d~=~6~cm (two 3~cm thick layers) as optimized for the J-PET
total-body prototype~\cite{MoskalKowalski}. The
two back-to-back 511~keV annihilation photons and the 1160~keV prompt photon were emitted assuming
that the activity is uniform along the 200~cm long line source positioned in the central axis of the
cylinder. For a single event both annihilation photons and prompt gamma were emitted from the same
point. The  direction of back-to-back photons and the direction of the prompt gamma were generated
independently, and in both cases the isotropic distribution was assumed. The diameter of the
cylinder was fixed to 80~cm whereas the length of the scanner was varied between 10~cm and 200~cm.

For a given axial field of view of the scanner the sensitivity $S$ was calculated as a function of the AFOV using a following formula:

\begin{multline}
  \int_{z=0}^{z=AFOV/2} dz \left[
    \int_{\theta_{min}(z)}^{\theta_{max}(z)} 
      (\epsilon_{det-a}(\theta) \cdot Att_a(\theta) )^2
      \cdot \sin\theta d\theta 
    \right] \\
  \left[
    \int_{\theta_{min}(z)}^{\theta_{max}(z)} 
    \epsilon_{det-p}(\theta) \cdot Att_p(\theta)
     \cdot \sin\theta d\theta 
    \right]
  \cdot \epsilon_{sel-a}^2 \cdot \epsilon_{sel-p} / L_{source},
    \label{eq::sensitivity}
\end{multline}
where $\theta$ denotes the  polar angle between the photon's direction of flight and the main axis of the tomograph. 
$L_{source} = 200~cm$ denotes the length of the annihilation line source. The detection efficiency
for annihilation photons $\epsilon_{det-a}$ and prompt gamma $\epsilon_{det-p}$ were calculated as $(1 -e^{-\mu \cdot d / sin\theta})$, where $\mu$ denotes the attenuation coefficient which 
for 511~keV (1160~keV) photons was assumed to be $\mu = 0.833~\mbox{cm}^{-1} (0.413~\mbox{cm}^{-1})$
and $\mu = 0.098~\mbox{cm}^{-1} (0.068~\mbox{cm}^{-1})$ for LYSO and plastic scintillators, respectively. $Att_a(\theta)$ and $Att_p(\theta)$ indicate attenuation in the body of annihilation photons and prompt gamma, respectively. The influence of attenuation was estimated as  
$e^{-\mu_{water} \cdot R_{phantom}/sin(\theta)}$ approximating the body as a 
cylindrical phantom with radius of $R_{phantom}~=~10~cm$ filled with water with  $\mu_{water}$ equal
to  $0.096~\mbox{cm}^{-1}$ and $0.066~\mbox{cm}^{-1}$ for 511~keV and 1160~keV photons, respectively. In the case
of the LYSO scintillators, the selection efficiency $\epsilon_{sel-a}$ of image forming signals for
511~keV photons was estimated as a fraction of the photoelectric effect which is equal to 0.34,
whereas for the plastic scintillators where 511~keV photons interact in practice only via Compton
effect $\epsilon_{sel-a}$ was set to 0.44~\cite{Moskal:2014sra} which results in a reduction of the scatter fraction to the level of  $\sim~35\%$~\cite{Kowalski2018}.
As regards the selection efficiency for the prompt photon with energy of 1160~keV which in both
cases (LYSO and plastic) is registered predominantly via Compton effect (in $\sim$ 100\% for
plastics and 89\% for  LYSO)  the selection efficiency $\epsilon_{sel-p}$ (both for LYSO
and plastic scintillators) was equal to 0.66, just as a fraction of the energy deposition spectrum with
deposited energy larger than the one from the 511~keV photons.

The above quoted values of attenuation coefficients and the fractions of photoelectric and Compton
effects were extracted from the data base maintained by the National Institute of Standards and
Technology~\cite{nist}.

\subsection{Detailed Monte Carlo simulations}
\label{sec::MC}
The detailed studies of the spatial and time resolution of the mean lifetime positronium images as a function of the time resolution of the detector system were performed on the example of the cost effective total-body PET built from plastic scintillators.  
Analogously as in reference~\cite{Moskal_2019_PMB} the ideal J-PET detector system consisting of
four concentric cylindrical layers, filled with plastic scintillator strips, was simulated. The
inner radius and axial field-of-view were chosen to be R~=~43~cm and AFOV~=~200~cm, respectively. For the single plastic strip, the cross section of  $7\times19$~mm$^2$  is chosen as it is used in the current version of the J-PET prototype~\cite{Szymon-Acta}.

In the plane comprising the central detector axis the 1-mm radius cylindrical $^{44}$Sc sources surrounded with different materials were simulated. 
Their positions were chosen according to the NEMA NU 2-2012 norm~\citep{NEMA:2012}, and
for each position a different mean lifetime of ortho-positronium was assumed, as it is listed in the
Table~\ref{tab:nema}. The values of the  mean ortho-positronium lifetime were chosen in the range
expected for ortho-positronium produced in the human body~\cite{Moskal_2019_PMB,JasinskaActaB:2017}. 
\begin{table}[hb]
  \centering
  \caption{Coordinates of simulated point-like sources positioned according to the NEMA norm. Each
  source is characterized by a different mean lifetime of ortho-positronium.}
  \label{tab:nema}
  \begin{tabular}{|C{0.2\textwidth}|C{0.3\textwidth}|C{0.3\textwidth}|} \hline
    Position & Coordinates [cm] & Simulated oPs mean lifetime [ns]\\ \hline
    1 & (1, 0, 0)     & 2.0  \\  
    2 & (10, 0, 0)     & 2.4 \\
    3 & (20, 0, 0)    & 2.8 \\
    4 & (1, 0, 75)     & 2.2 \\
    5 & (10, 0, 75)    & 2.6 \\
    6 & (20, 0, 75)    & 3.0 \\ \hline
  \end{tabular}
\end{table}

For the studies presented in this article $5\times 10^{6}$ events 
with the emission of prompt gamma from $^{44}Sc$ decay followed by the creation of ortho-positronium and its subsequent decay into two photons were generated. Subsequently, for each generated event, the response of the idealized J-PET scanner was simulated and then the standard $2\gamma$ image as well as the mean lifetime positronium image were reconstructed. 
Standard PET image showing the density distribution of annihilation points was performed using TOF
filtered back projection method  (TOF-FBP)~\cite{Conti2005} and the mean lifetime positronium image is reconstructed by the determination of the annihilation point and ortho-positronium lifetime on the event by event basis. Positronium lifetime image is constituted from mean ortho-positronium lifetimes determined for each image voxel.

The simulations were conducted in the following steps: 
For each event (i) the position of the annihilation was generated, in the arrangement required by
the  NEMA~\citep{NEMA:2012} norm (Table~\ref{tab:nema});
(ii) the prompt photon with energy of 1160~keV was emitted isotropically;
(iii) in each event a creation of ortho-positronium atom was assumed and its decay time was generated with the exponential probability density
distribution assuming mean lifetime depending on the position (see Table~\ref{tab:nema});
(iv) for each event an ortho-positronium annihilation (pick-off or conversion processes) into two 511~keV back-to-back photons was generated assuming the isotropic emission independent of the emission of the prompt gamma;  
(v) hit-positions and energy depositions of annihilation photons and prompt gamma in the J-PET scanner were simulated taking into account cross sections for Compton interactions of gamma photons in plastic scintillators;
(vi) the experimental hit-position resolutions (axial and radial) was accounted for by smearing the generated positions with Gaussian functions having 5~mm (FWHM) as expected for J-PET scanner with the assumed SiPM and WLS readout~\cite{Kowalski2018,Moskal_2019_PMB};
(vii) the time resolution was included by smearing the generated interaction times with Gaussian
distributions corresponding to CRT values of $\{140, 50, 10\}$ ps (the values after smearing will be referred to as 
{\it registered});
(viii) for each {\it registered} event,  the time and position of annihilation point as well as the time of the emission of prompt gamma was reconstructed; and finally (ix) for each voxel of the image a mean  ortho-positronium lifetime was reconstructed.

The simulation methods for the above listed steps 
were described in details in the previous publications~\citep{Moskal_2019_PMB,Moskal:2016ztv,Kaminska:2016fsn,Moskal:2014sra,Moskal:2014rja}.

\section{Results}
\label{sec::results}

\begin{figure}[h]
  \centering
    \includegraphics[width=0.5\textwidth]{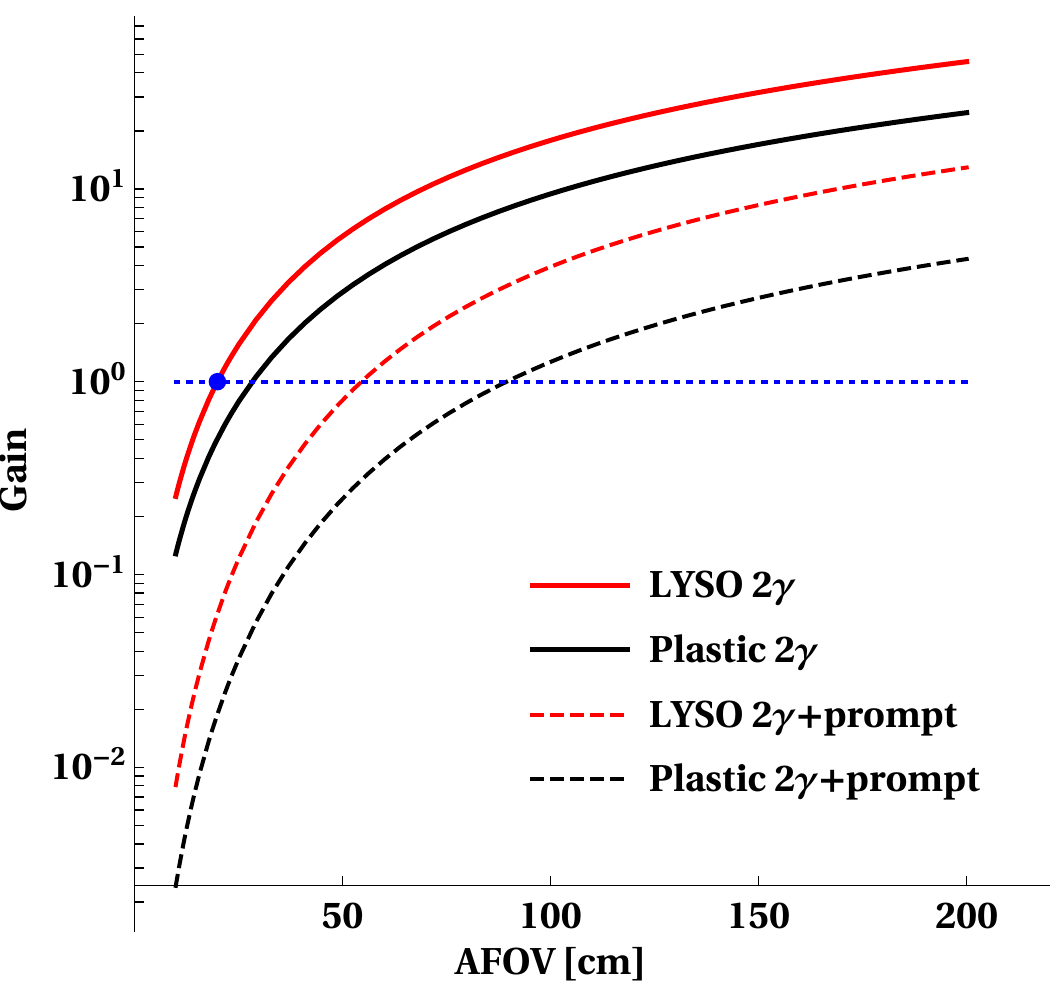}
    \caption{ Relative gain in sensitivity $S$ as a function of the AFOV of the scanner. The gain is calculated relative to sensitivity for the standard $2\gamma$ imaging using LYSO PET with AFOV~=~20~cm (blue dot). The gains for $2\gamma$ (solid lines) as well as for $2\gamma + \gamma_{prompt}$ (dashed lines) are shown for LYSO and plastic scintillators as indicated in the legend. The red and black colour indicate result for the LYSO and plastic scintillators, respectively.
    \label{fig::gain_sensitivity}
    } 
\end{figure}

Figure~\ref{fig::gain_sensitivity} presents the gain of the sensitivity 
calculated as defined by (Eq.~\ref{eq::sensitivity}) in section~\ref{sec::materials} The gain is calculated with respect to the
sensitivity of the PET from LYSO crystals with AFOV~=~20~cm for the standard $2\gamma$ imaging (indicated as blue dot in
the figure). The relative gain is presented as a function of AFOV of the scanner.  Dashed lines
indicate that for the ($2\gamma + \gamma_{prompt}$) triple coincidences with AFOV~=~200~cm  the
achievable gain is equal to factor of about  12.2 and 4.7 for LYSO and plastic based PET, respectively.
It indicates that the $2\gamma \ + \gamma_{prompt}$ positronium imaging is feasible with the total-body PET scanners with sensitivities 
even higher than the one of the current PET with AFOV~$\sim$~20~cm. 
For the comparison, in order to cross-check the estimations, the gain for the standard $2\gamma$
imaging is shown. The red and black solid lines indicate sensitivity gain expected for the
registration of the image forming events with two back-to-back annihilation photons ($2\gamma$)
calculated for the LYSO and plastic based scanners, respectively. As
expected,~\cite{Cherry2017,Cherry2018} for the whole body scan with total-body LYSO PET the
sensitivity is increased by more than factor of forty. The black solid line indicates that for the
whole-body scan, the total-body PET from plastic scintillators also provides a~large, about twentyfold, gain in sensitivity.

\begin{figure}[h]
  \centering
    \includegraphics[width=0.6\textwidth]{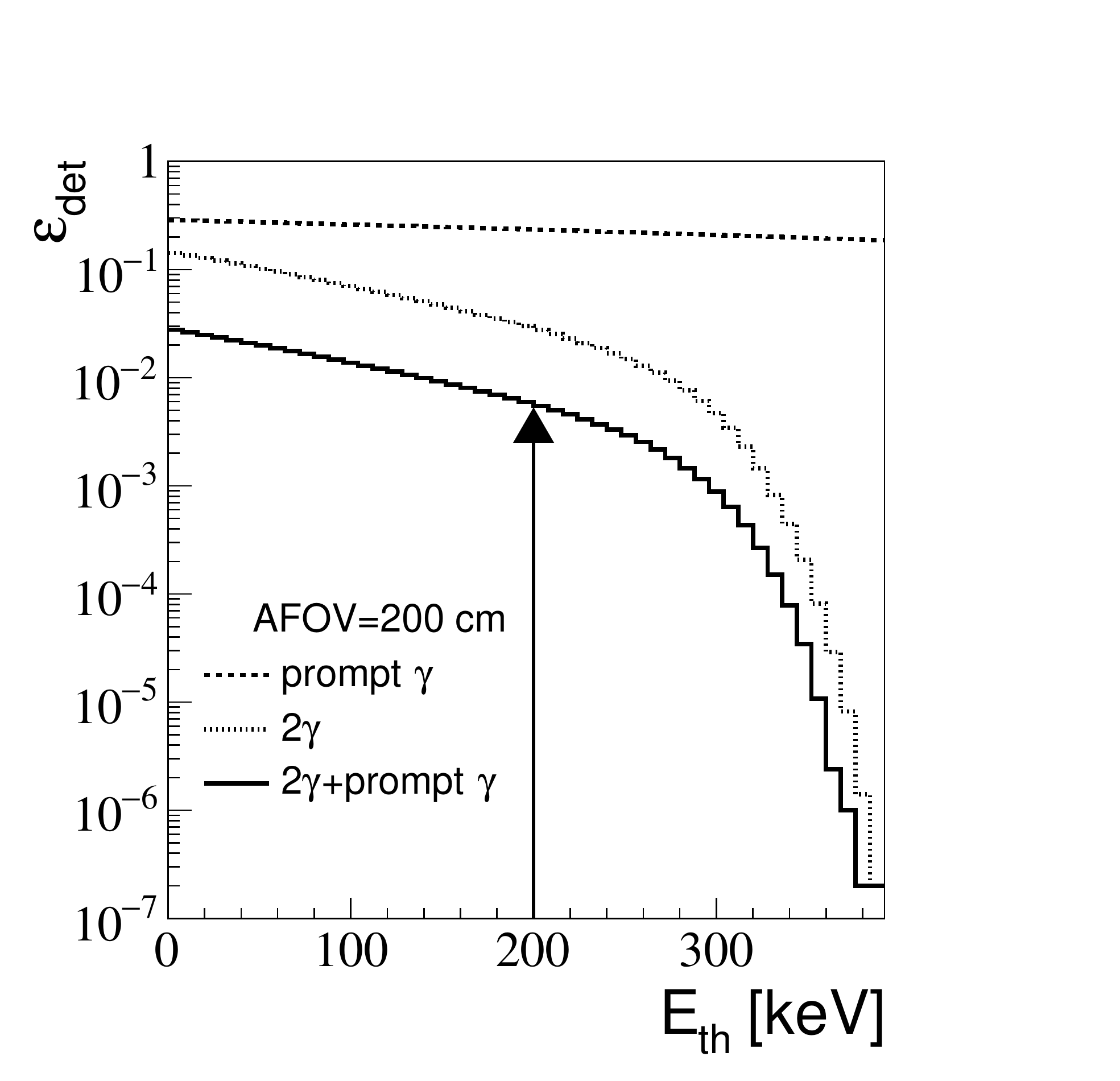}
    \caption{     Registration efficiency (taking into account geometrical acceptance, probability of gamma quanta 
    registration in the plastic scintillator and \mbox{J-PET} resolution) as a function 
    of applied threshold for the cases of prompt gamma (dashed line),  two back-to-back 511~keV photons (dotted line)  and two photons simultaneously with the prompt gamma with energy loss higher than 400~keV (solid line).
    \label{fig::efficiency2}
    } 
\end{figure}
Figure~\ref{fig::efficiency2} presents sensitivities (detection efficiency) calculated
using detailed Monte-Carlo method (section \ref{sec::materials}) for the case of the total-body plastic PET
with AFOV~=~200~cm and total plastic thickness of ~6~cm. The dashed line indicates overall
geometrical and detection efficiency for the registration of the prompt gamma (1160 keV) and the
dotted histogram shows detection efficiency for the registration of the back-to-back 511~keV
photons. The results are shown as a function of the energy deposition threshold. Solid line
indicates overall efficiency for the registration of back-to-back photons with coincident detection
of prompt photon with energy deposition larger than 400~keV. Such condition enables to disentangle
prompt gamma from 511~keV photons. In order to decrease the scatter fraction down to 35\% a minimum
energy deposition of 200~keV is required for the 511~keV
photons~\cite{Kowalski2018,Kowalski2016,Moskal:2014rja}. Therefore, for the
$2\gamma$ positronium mean lifetime imaging of the 200~cm long objects, with the plastic total-body
PET, a total detection and selection efficiency of about 0.5\% is expected (as indicated in
Fig~\ref{fig::efficiency2} by an arrow). 
Thus with the plastic total-body PET about $6.6~\cdot~10^8$ image forming triple coincidence events
($2\gamma + \gamma_{prompt}$) may be collected assuming (i) 20~min scan according to the standard
whole-body protocol, (ii) activity of 370 MBq (10 mCi) administered to the patient~\cite{Cherry2018}, (iii)
detection sensitivity of $0.5\%$, and (iv) 30\% fraction of positron annihilations through the
formation of ortho-positronium atoms in the body. 
This corresponds to about $10^4$ event forming events per cubic centimeter for the plastic based
total-body PET. While, when taking into account ratio between dashed lines in
Fig.~\ref{fig::gain_sensitivity}, $2.6\cdot 10^4$ events per cubic centimeter are expected for LYSO based total-body PET. These numbers will be used later on to estimate mean lifetime resolution for positronium imaging. 

\begin{figure}[htbp]
  \centering
    \begin{subfigure}{\textwidth}
      \includegraphics[width=0.34\textwidth, height=0.34\textwidth, keepaspectratio=true, trim={0.5cm 0.5cm 1.5cm 1.5cm}, clip]{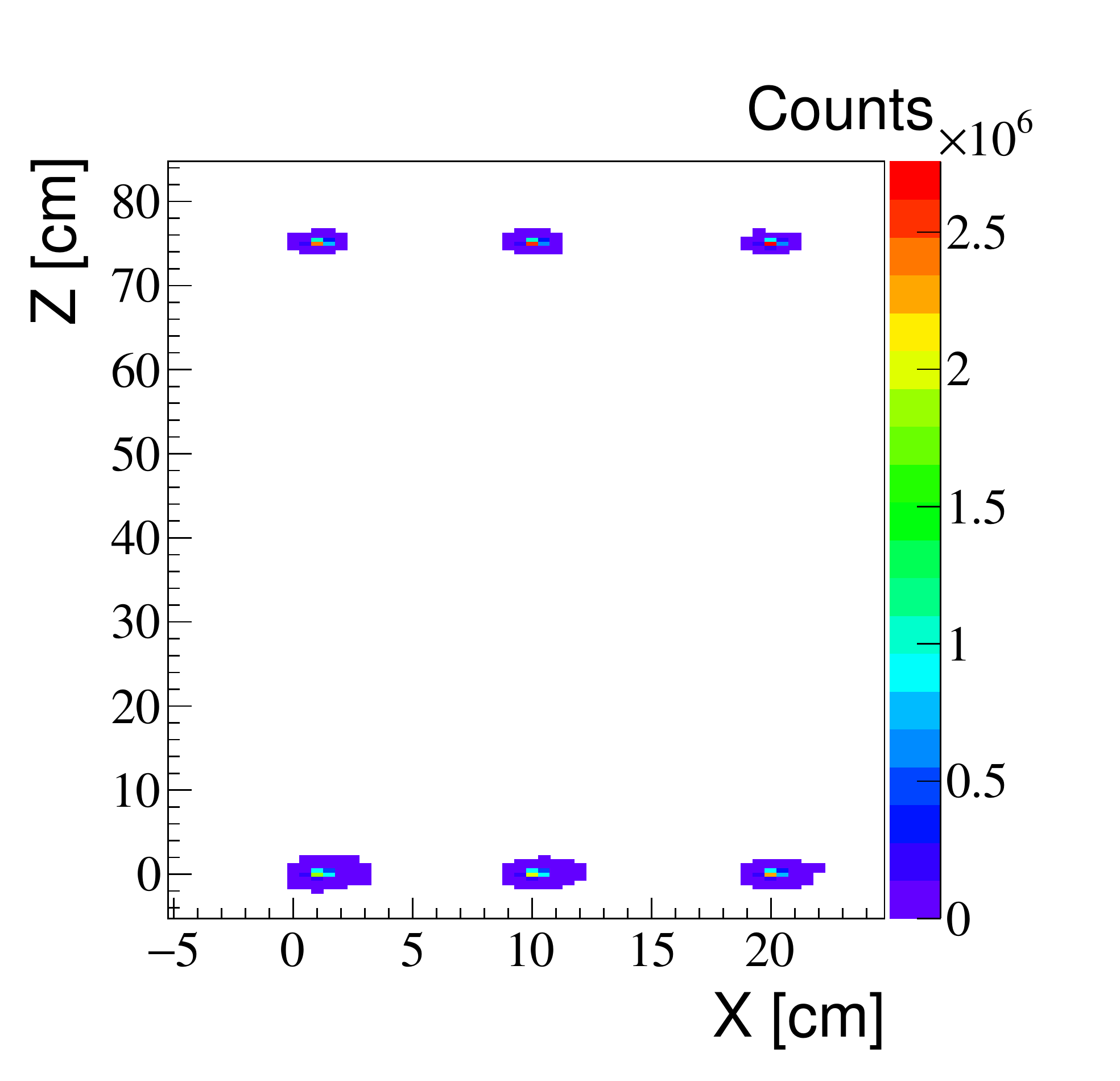} 
      \includegraphics[width=0.4\textwidth, height=0.4\textwidth, keepaspectratio=true, trim={0.2cm 0.2cm 0.3cm 0.7cm}, clip]{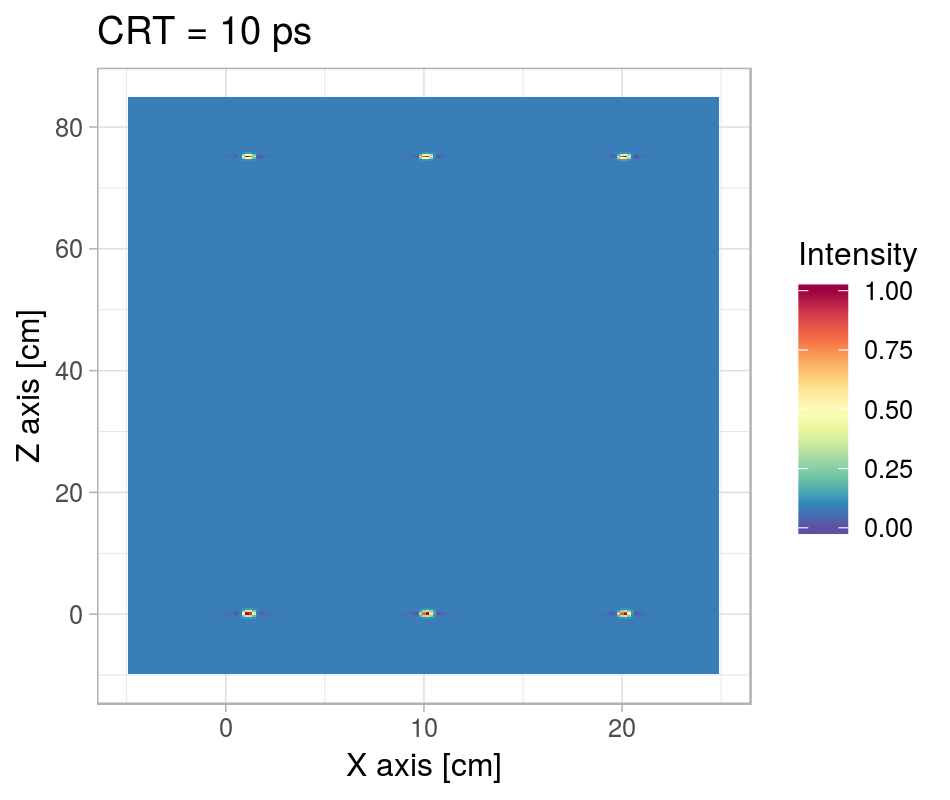}
      \includegraphics[width=0.24\textwidth,  keepaspectratio=true, trim={0.2cm -1.5cm 2.3cm 0.7cm}, clip]{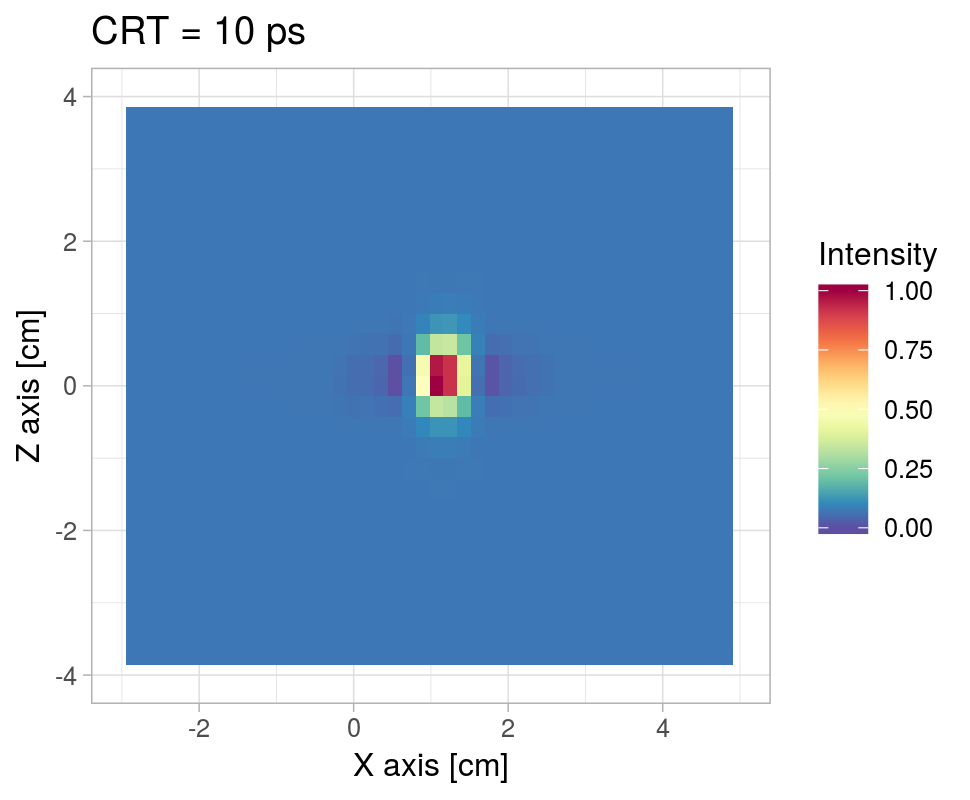}
      \label{fig::entries_3}
    \end{subfigure}
    \begin{subfigure}{\textwidth}
      \includegraphics[width=0.34\textwidth, height=0.34\textwidth, keepaspectratio=true, trim={0.5cm 0.5cm 1.5cm 1.5cm}, clip]{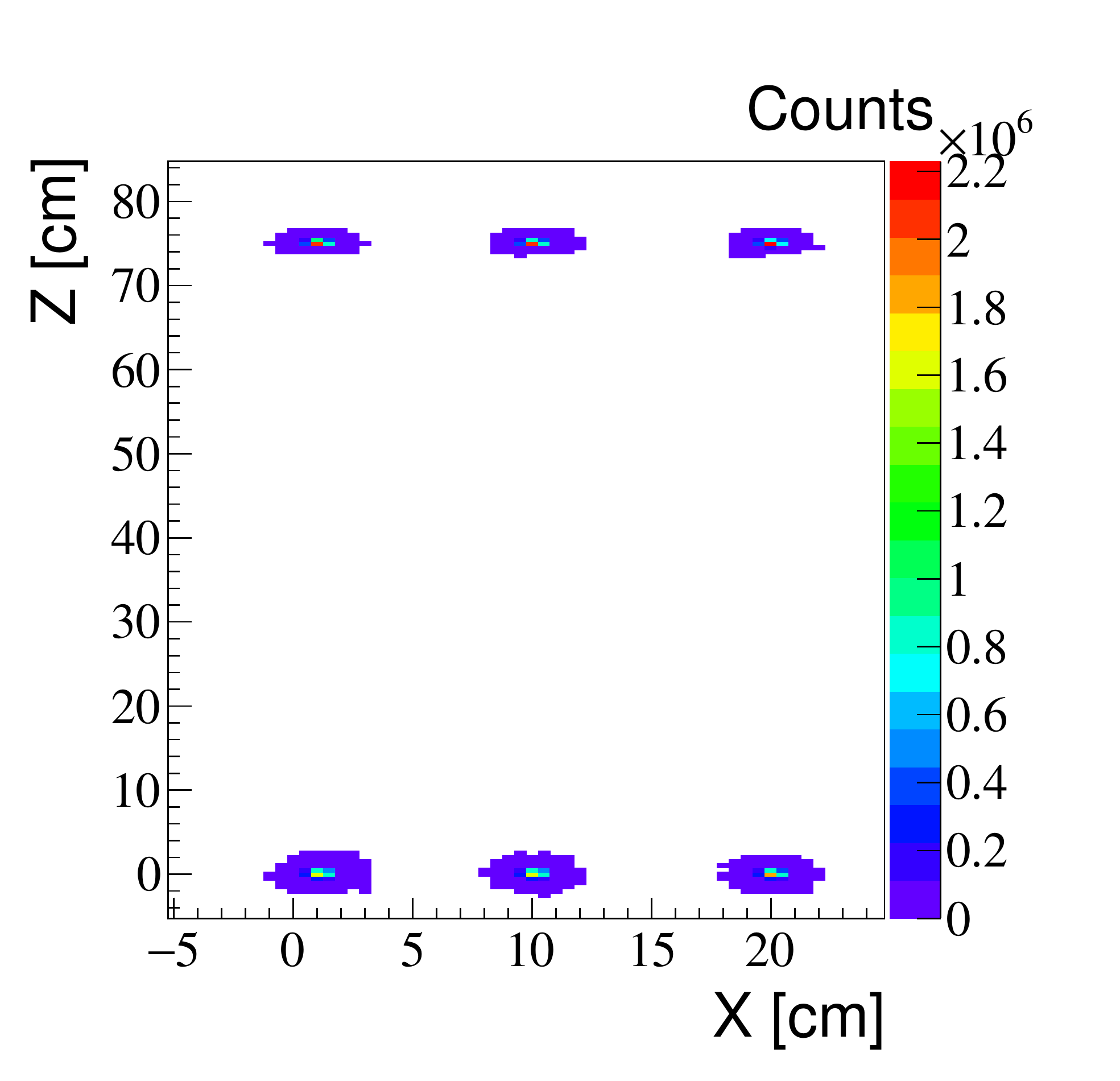} 
      \includegraphics[width=0.4\textwidth, height=0.4\textwidth, keepaspectratio=true, trim={0.2cm 0.2cm 0.3cm 0.7cm}, clip]{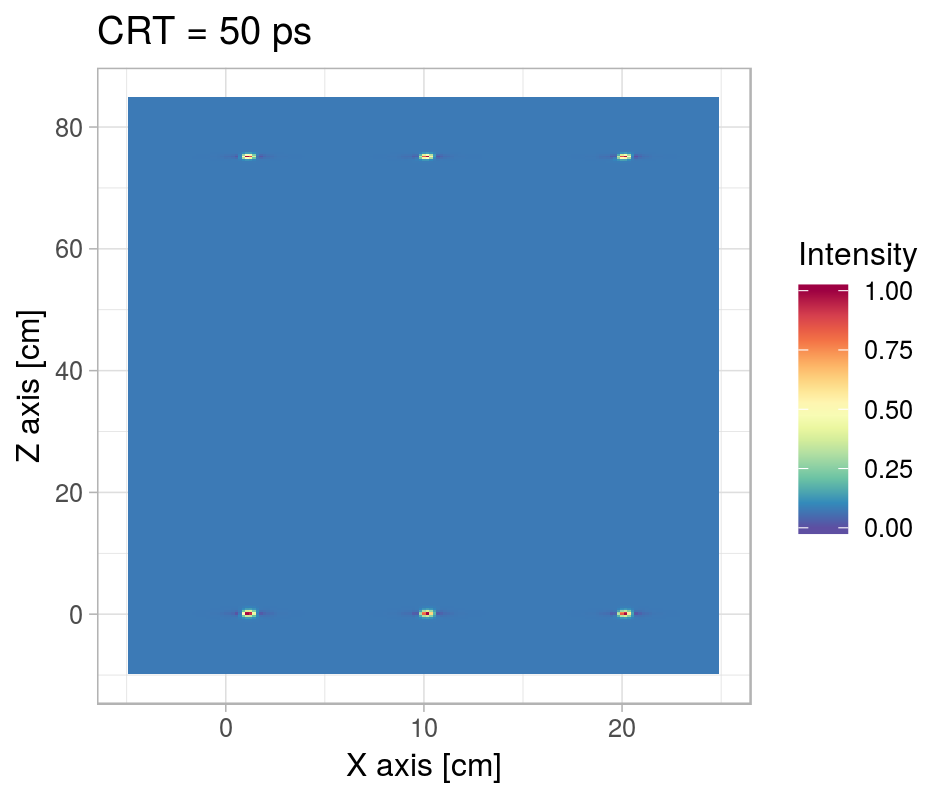}
      \includegraphics[width=0.24\textwidth,  keepaspectratio=true, trim={0.2cm -1.5cm 2.3cm 0.7cm}, clip]{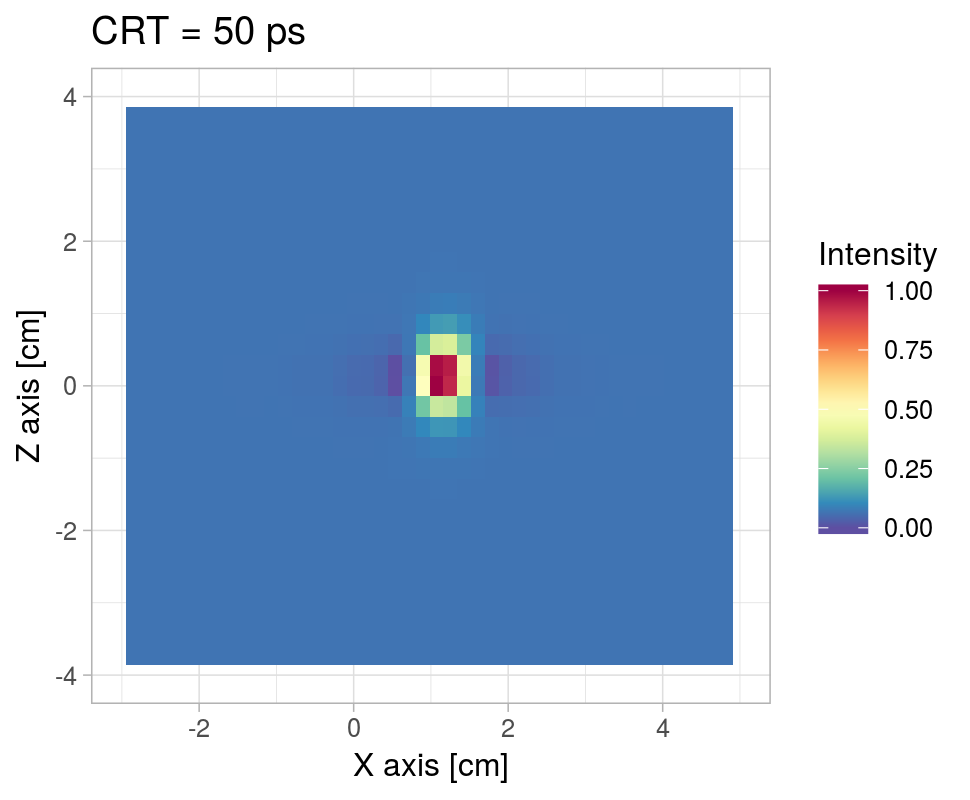}
      \label{fig::entries_15}
    \end{subfigure}
    \begin{subfigure}{\textwidth}
      \includegraphics[width=0.34\textwidth, height=0.34\textwidth, keepaspectratio=true, trim={0.5cm 0.5cm 1.5cm 1.5cm}, clip]{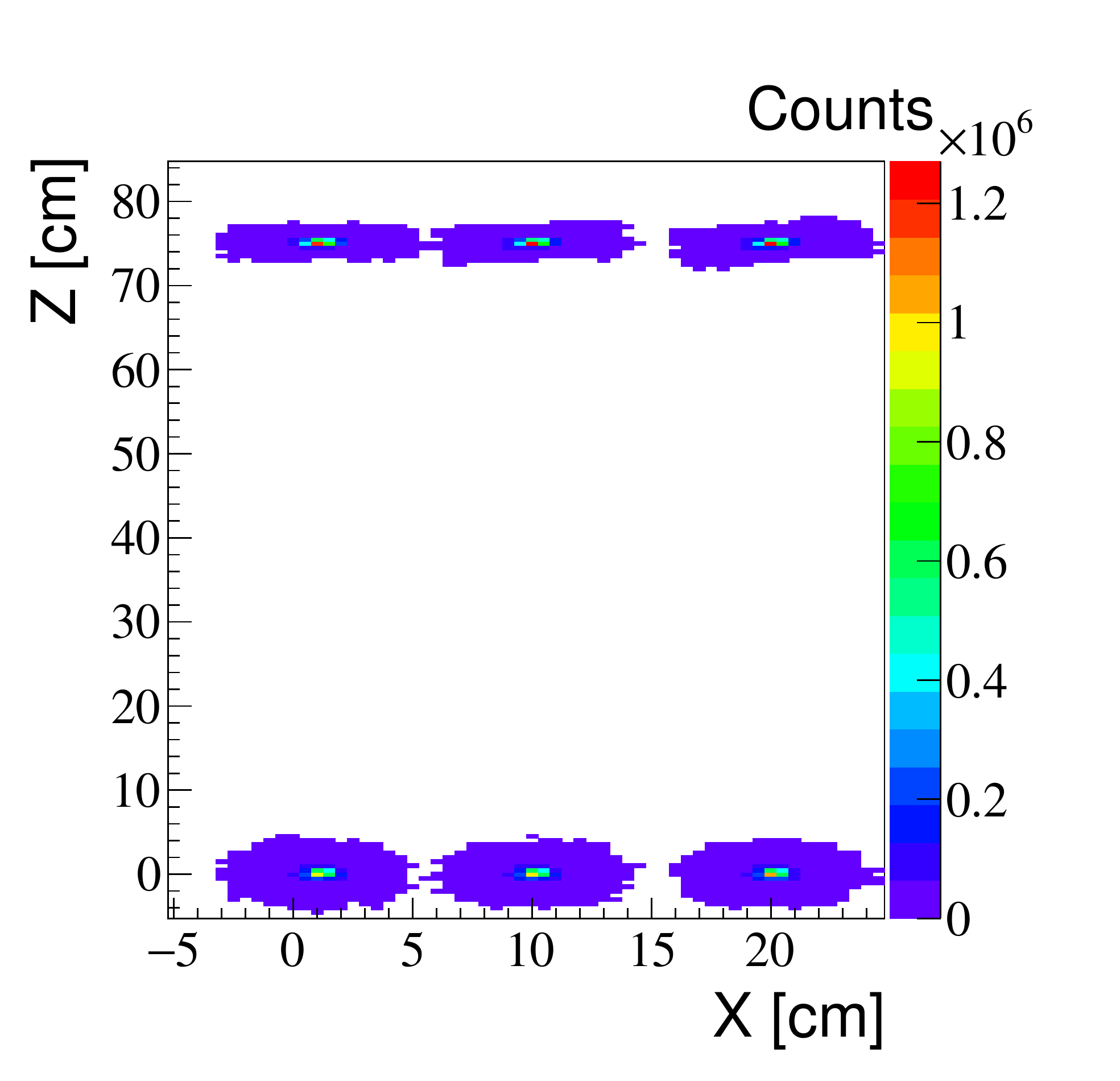} 
      \includegraphics[width=0.4\textwidth, height=0.4\textwidth, keepaspectratio=true, trim={0.2cm 0.2cm 0.3cm 0.7cm}, clip]{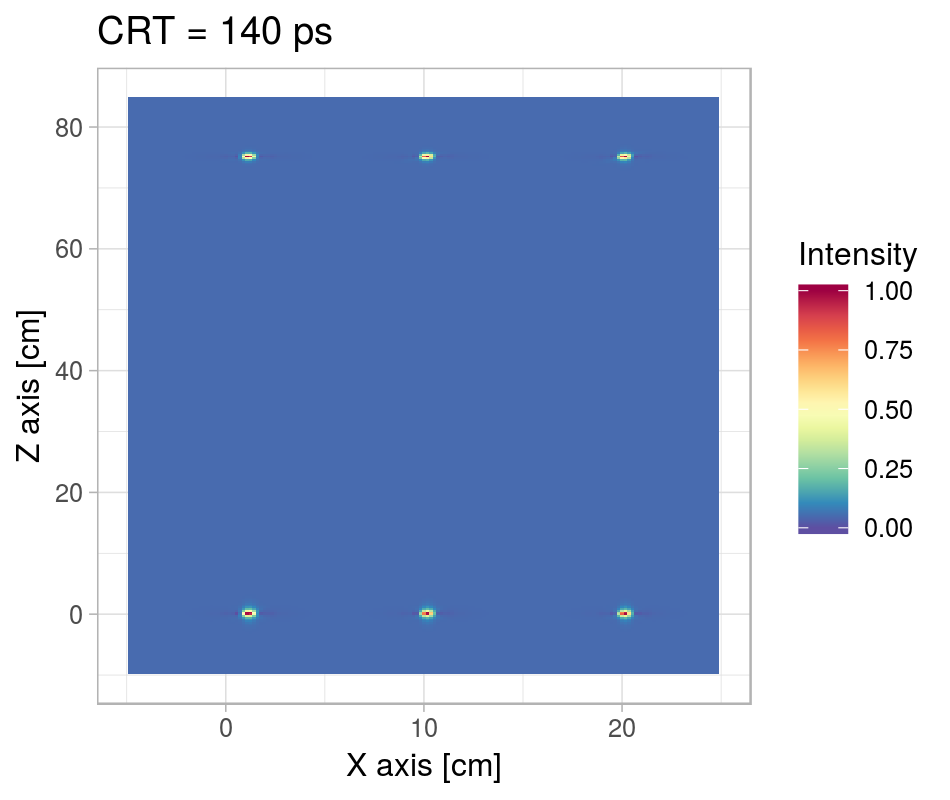}
      \includegraphics[width=0.24\textwidth,  keepaspectratio=true, trim={0.2cm -1.5cm 2.3cm 0.7cm}, clip]{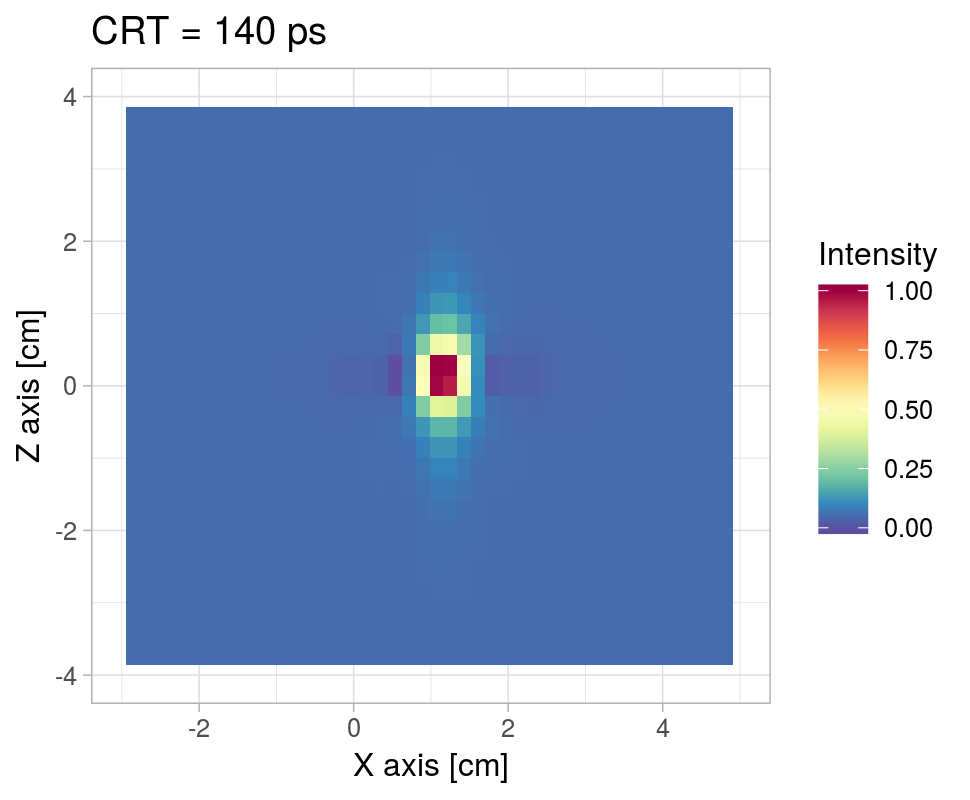}
      \label{fig::entries_42}
    \end{subfigure}
    \caption{\Left Reconstructed distribution of annihilation point spatial coordinates.
    The voxel size is equal to $5\times5\times5~\mbox{mm}^3$. 
    \Middle Reconstructed image of six sources obtained while applying the TOF-FBP algorithm.
    The voxels:  $1.8\times1.8\times2.9~\mbox{mm}^3$.
    \Right Reconstructed source placed at $(x,y,z)=(1, 0, 0)$~cm.    
    Each row show results with different resolution: CRT$=10$~ps (top), CRT$=50$~ps (middle)
    and CRT$=140$~ps (bottom).
      }
    \label{fig:positionRec}
\end{figure}

Fig.~\ref{fig:positionRec} compares standard PET images determined for the point-like sources arranged according to the NEMA norm recommendations~(Tab.~\ref{tab:nema}). The coronal $XZ$ cross-section along $y=0$~cm are shown.
Left column shows images obtained directly as a density distribution of annihilation points assuming TOF resolution of CRT~=~10~ps (top row), CRT~=~50~ps (middle row), and CRT~=~140~ps (bottom row). Middle column indicates images determined using the TOF filtered back projection (TOF-FBP) algorithm,
and right column presents enhanced view of images of the source close to the centre of the scanner.
The obtained point spread functions (PSF) are listed in Tab.~\ref{tab:PSF}. The result indicates that already for CRT~=~50~ps the PSF of the direct image is comparable with the PSF achievable when applying reconstruction with TOF-FBP algorithm.

\begin{figure}[htbp]
  \centering
    \begin{subfigure}{.45\textwidth}
      \includegraphics[width=\textwidth]{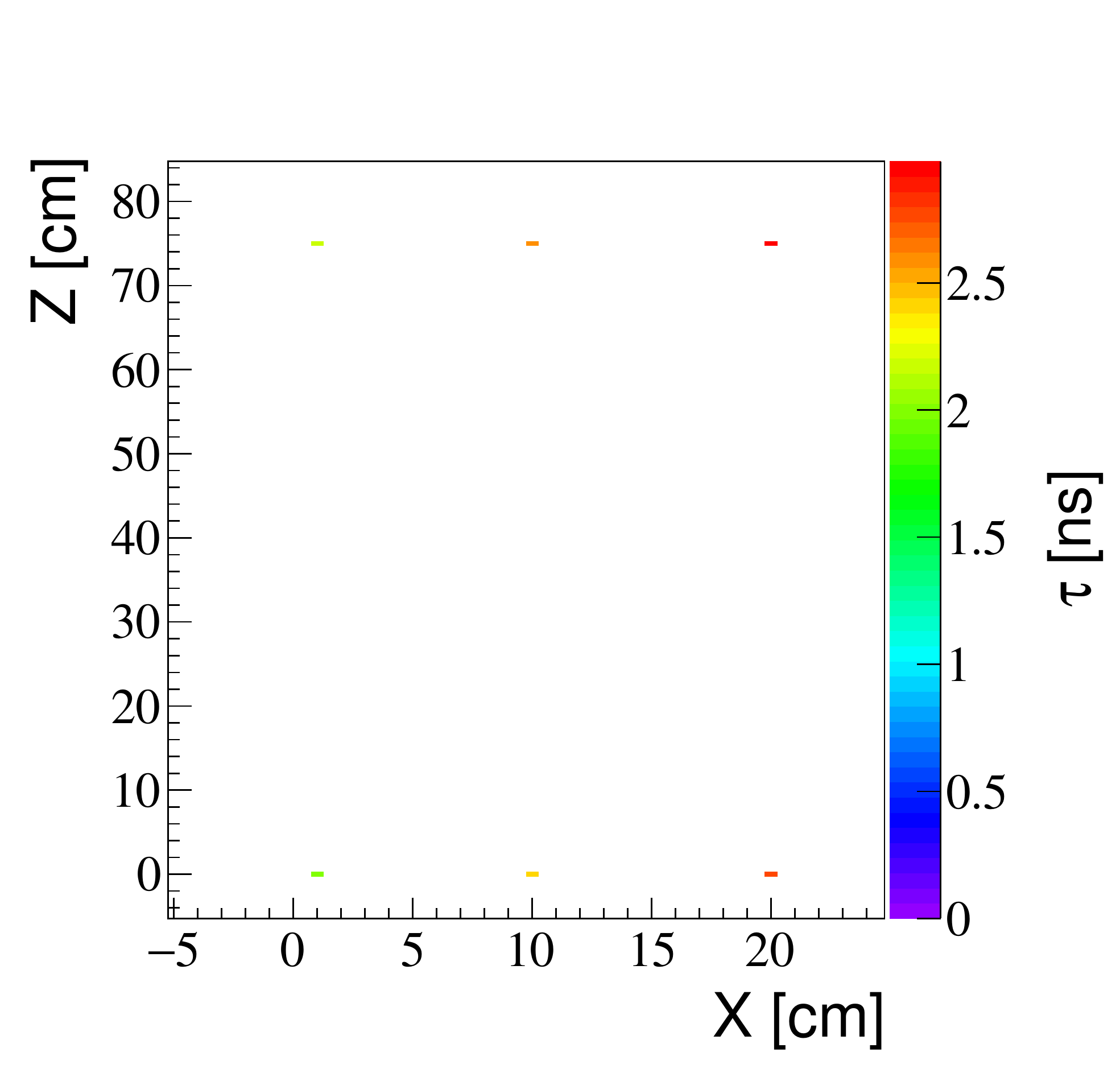}
      \caption{Generated positronium lifetime distribution.}
      \label{fig::lifetime_gen}
    \end{subfigure}
    \begin{subfigure}{.45\textwidth}
      \includegraphics[width=\textwidth]{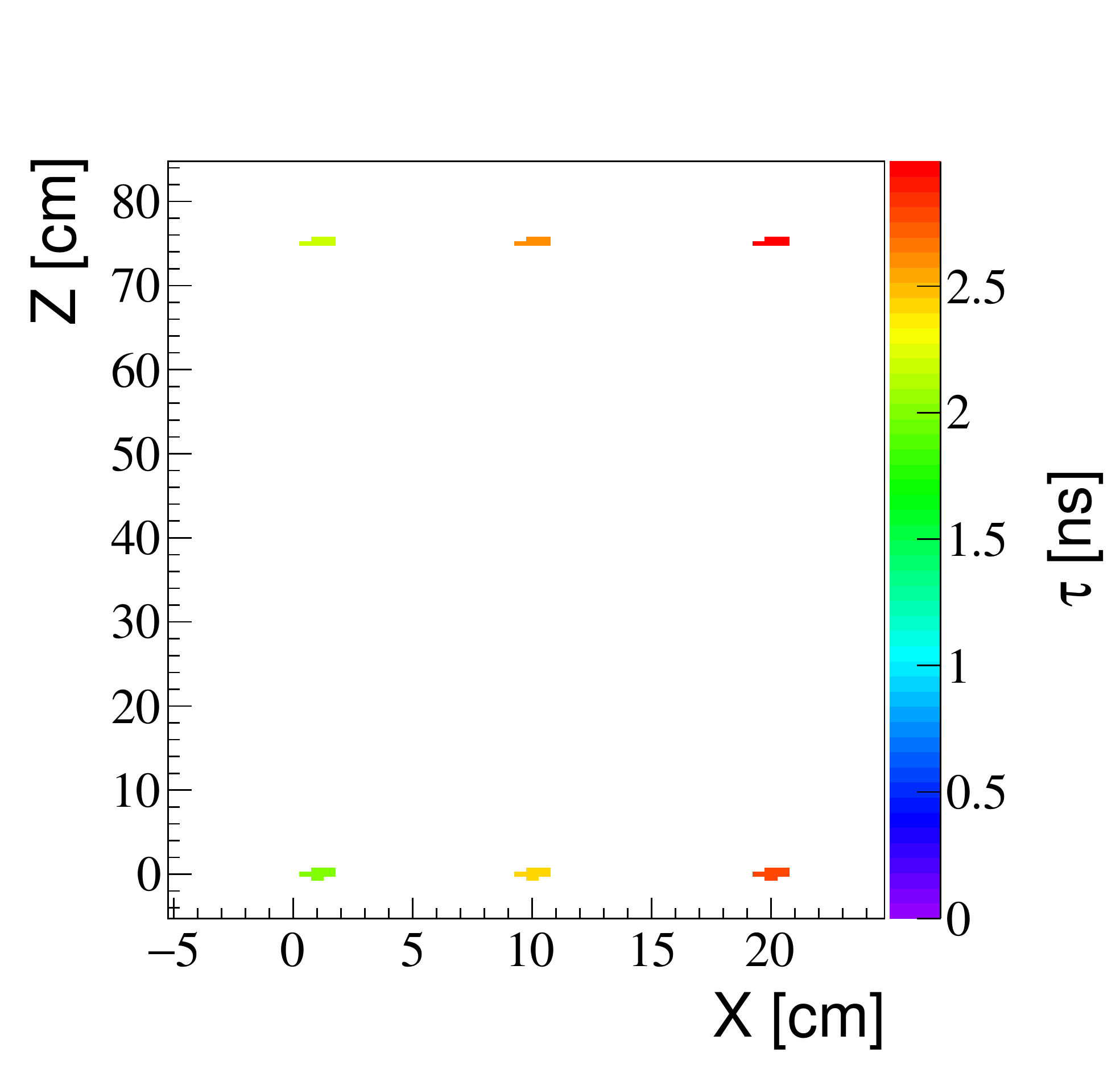}
      \caption{Reconstructed positronium lifetime distribution (CRT=10~ps).}
      \label{fig::lifetime_3}
    \end{subfigure}
    \begin{subfigure}{.45\textwidth}
      \includegraphics[width=\textwidth]{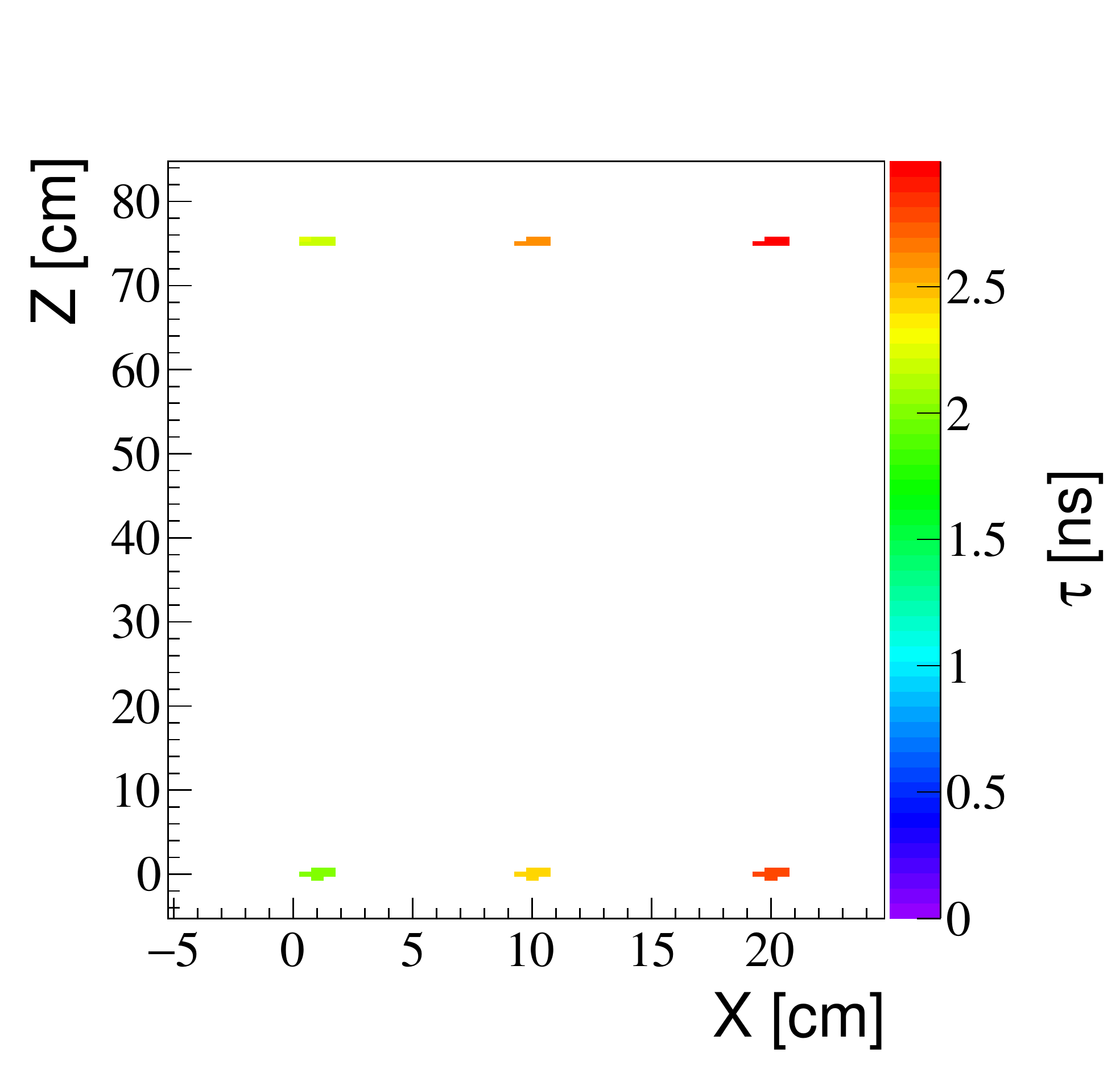}
      \caption{Reconstructed positronium lifetime distribution (CRT=50~ps).}
      \label{fig::lifetime_15}
    \end{subfigure}
        \begin{subfigure}{.45\textwidth}
          \includegraphics[width=\textwidth]{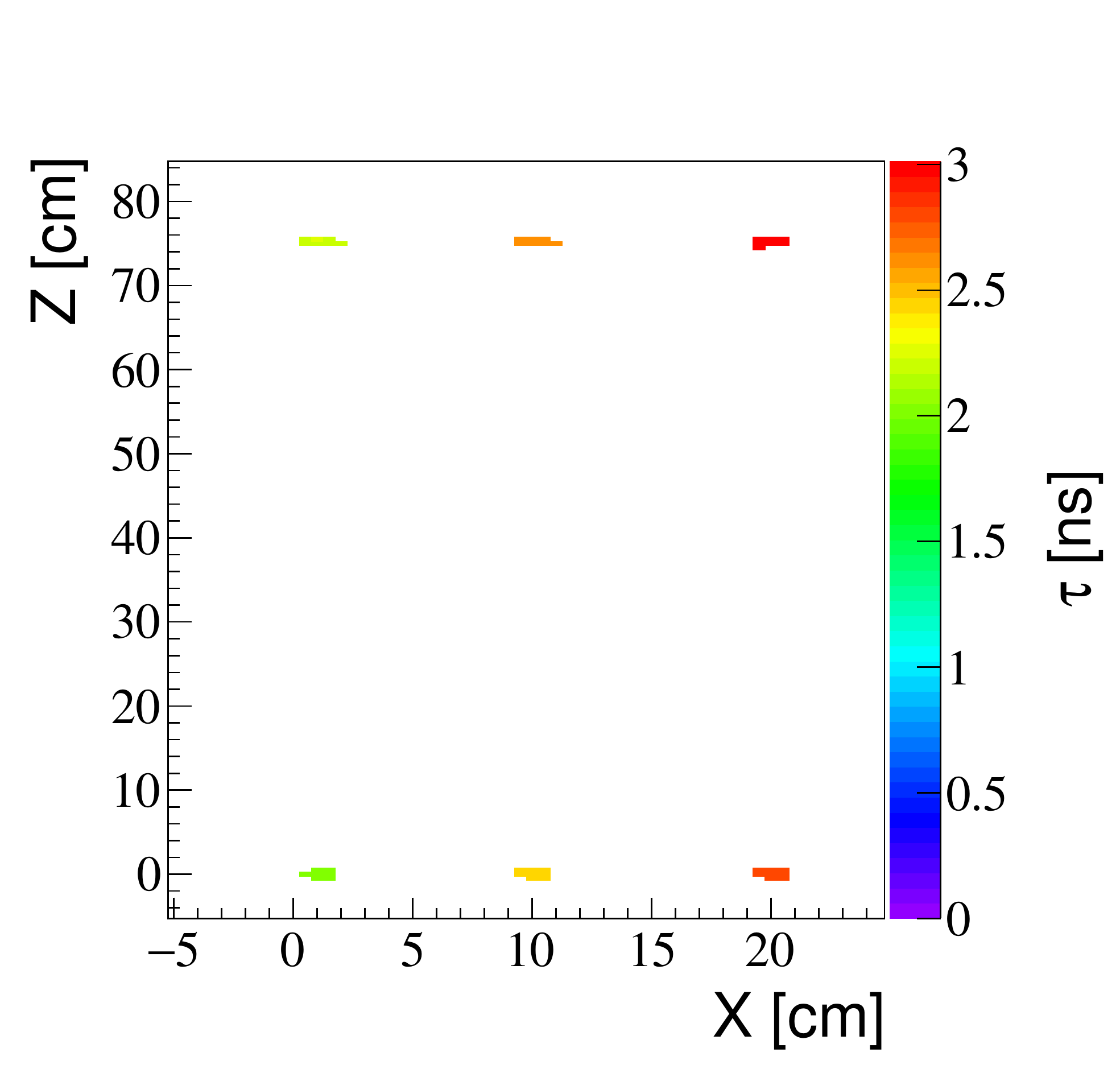}
            \caption{Reconstructed positronium lifetime distribution (CRT=140~ps).}
          \label{fig::lifetime_42}
        \end{subfigure}
    \caption{
      Distributions of generated positronium lifetimes~(\subref{fig::lifetime_gen}),
      and reconstructed ones, assuming the CRT value of
      10~ps (\subref{fig::lifetime_3}), 50~ps (\subref{fig::lifetime_15}) and  140~ps (\subref{fig::lifetime_42}).
      The voxel size is equal to $5\times5\times5~\mbox{mm}^3$. 
    }
    \label{fig:lifetimeRec}
\end{figure}

\begin{figure}[h]
  \centering
  \includegraphics[width=0.45\textwidth]{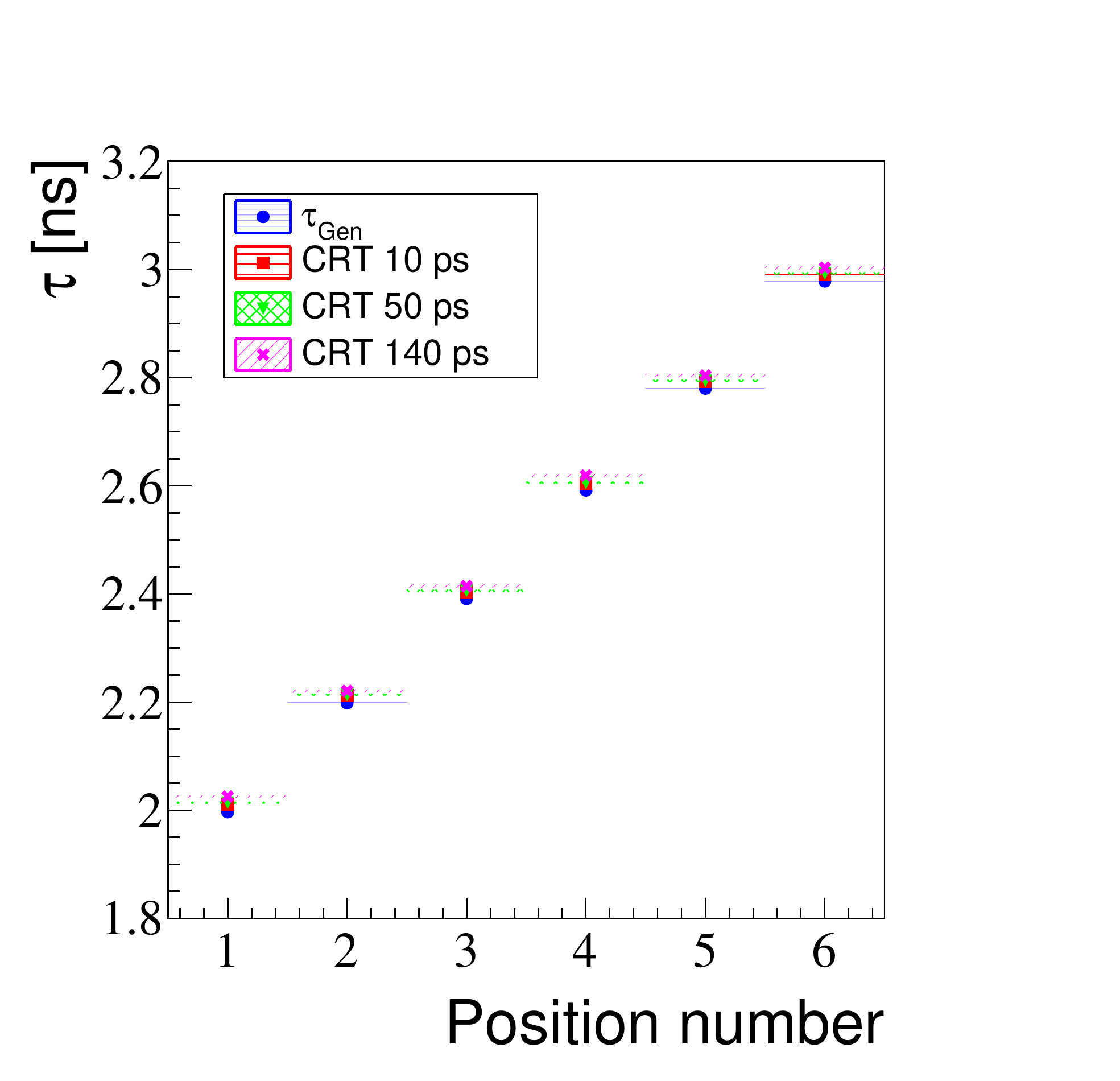}
    \includegraphics[width=0.45\textwidth]{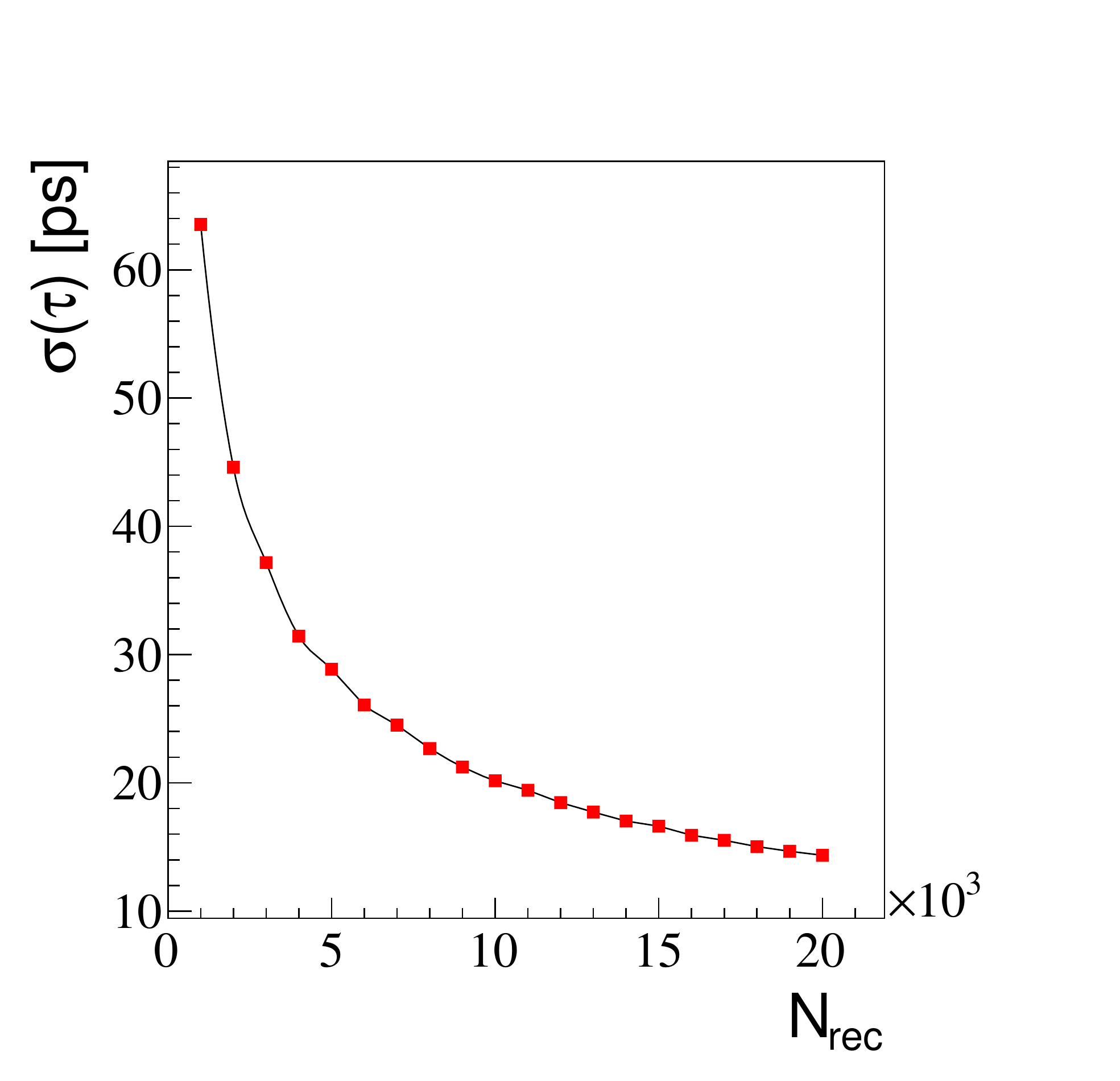}
    \caption{
      \Left
    Comparison between generated positronium mean lifetime and reconstructed one assuming
    different detector CRT resolutions as a function of NEMA position number 
    (see~Table~\ref{tab:nema}). Differences between obtained results are in the order of
    $\mathcal{O}$(10~ps).
    \Right
    Resolution of the mean lifetime determination as a function of detected entries in a single voxel.
        \label{fig::pointsComparison}
      }
\end{figure}

Finally the results for the mean lifetime positronium image are presented. 
The triple coincidence events ($2\gamma + \gamma_{prompt}$) were simulated and analyzed as it is described in section~\ref{sec::materials}. 
When using events with triple coincidences $2\gamma + \gamma_{prompt}$ it is possible to reconstruct position and time of the creation of two back-to-back photons and the time of the creation of the positron. The latter can be deduced based on the time and position of interaction of the prompt  gamma and the position of the annihilation point. Thus for each voxel of the image a spectrum of differences between the time of annihilation and the time of positron emission can be created enabling  mean lifetime ortho-positronium reconstruction.  
Fig.~\ref{fig:lifetimeRec} compares ortho-positronium mean lifetimes images reconstructed based on simulated signals (described in section~\ref{sec::materials}) to the generated image. The simulation and reconstruction was performed for three cases assuming time resolution of the scanner to be CRT~=~10~ps, 50~ps and 140~ps. All reconstructed images reflect the generated image very well. The quantitative comparison is shown in the left panel of Fig.~\ref{fig::pointsComparison}. Figure shows that for each simulated source position, the reconstructed mean lifetimes agree with the generated values within about 10~ps. 
The resolution of the reconstructed mean lifetime depends on (i) the detector's time and spatial
resolution, (ii) on the mean ortho-positronium lifetime ($\tau_{tissue}$), as well as (iii) on the
number of events assigned to a given voxel. The right panel of Fig.~~\ref{fig::pointsComparison}
presents the resolution as a function of the number of events.  The result indicates that~a resolution of better than 20~ps (RMS) is achievable for the number of image forming events ($10^4$) expected per cubic centimeter in a 20 minutes  total-body scan. The result is valid for all tested CRT values since they are more than order of magnitude smaller than the mean lifetime of ortho-positronium atoms in the body (few nanoseconds), and therefore the reconstruction resolution of the mean lifetime depends predominantly on  $\tau_{tissue}$ and may be approximately estimated as $\tau_{tissue}/\sqrt{N}$, where N denotes number of events in a given voxel. It is important to stress that in Fig~\ref{fig::efficiency2} the values of sensitivities do not include losses in the number of events due to attenuation and scattering of photons in the patient. Inclusion of attenuation would decrease the number of positronium-image forming events by about an order of magnitude. Therefore, at the present early stage of the development of this method it would be appropriate to strive to achieve first experimental images with larger voxel's size (e.g. $2~\times~2~\times~2~cm^3$) 
ensuring the statistics of ($10^4$) events per voxel.

\begin{table}[hbp]
  \centering
    \begin{tabular}{|C{0.1\textwidth}|C{0.15\textwidth}|C{0.15\textwidth}|C{0.15\textwidth}|C{0.15\textwidth}|} \hline
      &  \multicolumn{2}{c|}{PSF direct image [mm]} &  \multicolumn{2}{c|}{PSF TOF-FBP image [mm]}   \\
         CRT [ps] &  radial  & axial   & radial  &  axial  \\ \hline
         10 &  4.2 & 4.8 & 4.5 & 5.5\\ \hline
         50 &  5.4 & 5.2 & 4.5 & 5.5 \\ \hline
          140 & 10.2 & 6.8 & 5.0 & 6.0 \\ \hline
    \end{tabular}
    \caption{Point Spread Function (PSF) obtained for the direct and TOF-FBP images
             as a function of CRT. Values presented are for position  $(x=20, y=0, z=75)$~cm.
    \label{tab:PSF}
    }
\end{table}

\section{Discussion}
\label{sec::Discussion}
Recently it was reported that the mean lifetime of ortho-positronium, which is copiously produced in
the body during the PET investigation, may be used as the in-vivo diagnostic indicator complementary
to the currently available SUV index~\cite{Moskal_2019_PMB,Moskal2019NatureReview}. In
this article we assess the feasibility of ortho-positronium mean lifetime imaging using $2\gamma$
decays which occurs in the tissue due to the pick-off and ortho- to para-positronium conversion
processes. We have shown that with the advent of the total-body PET and the improvement of time
resolution to tens of picoseconds the $2\gamma$ ortho-positronium mean lifetime imaging may be
performed with the sensitivity and spatial resolution comparable with the current standard $2\gamma$
metabolic imaging obtainable by PET scanners with AFOV of about 20~cm. We also have shown that
having time resolution of tens of picoseconds the $2\gamma$ mean lifetime positronium imaging
becomes feasible for the cost effective total-body J-PET scanner built from plastic scintillators.
Calculations were preformed taking into account $^{44}Sc$ labeled radiopharmaceutical emitting prompt gamma with the energy of 1160~keV. In the calculations of the sensitivity gain the attenuation of both two 511~keV annihilation and prompt photons in the 20~cm diameter water phantom were taken into account. Quantitative estimations show that for the whole-body scan, the overall sensitivity for registration and image forming selection of 
$2\gamma + \gamma_{prompt}$ 
events exceeds the sensitivity of current PET $2\gamma$ metabolic imaging twelvefold in case of
total-body PET based on LYSO scintillators and fourfold in case of the total-body PET from plastic
scintillators. Moreover, the obtained result (Fig.~\ref{fig::gain_sensitivity}) indicates that the sensitivity for ortho-positronium mean
lifetime image is becoming comparable with current PET sensitivities  for AFOV of about 56~cm (LYSO
PET) and  87~cm (plastic PET). Finally the mean lifetime resolution achievable with the
presented method for total-body PET is ranging between 10~ps and 20~ps for LYSO and plastic PET
systems and it is predominantly  due to the large (few ns) ortho-positronium mean lifetime while it
is fairly independent of the more than order of magnitude smaller CRT values.    

Positronium mean lifetime imaging with two photons, discussed in this article, is beneficial in
relation to imaging using three photons due to the smaller attenuation of photons in the body and
higher detection sensitivity. In  case of the ortho-positronium $2\gamma$ imaging  the attenuation
of photons in the body will be much smaller than for the $oPs \to 3\gamma$ imaging. This is because
in the latter case there is one more photon which needs to escape from the body and the energy of
photons in case of $oPs \to 3\gamma$ ranges from 0 to 511~keV and thus these photons having energy
lower than 511~keV are on the average more strongly absorbed in the body with respect to 511~keV
photons from the $2\gamma$ annihilations. Moreover, the ortho-positronium mean lifetime imaging based on the $oPs \to 2\gamma$  events may directly be applied in the present TOF-PET systems. Making the application of the proposed method realistic in the near future. 

\section{Conclusions}
\label{sec::Conclusions}

The method discussed in this article, in general enables to determine a spectrum of positron
lifetime on a voxel by voxel basis. Allowing to determine not only an image of the mean lifetime of
ortho-positronium (or in general a distribution of mean ortho-positronium
lifetimes~\cite{Dulski2017}), but it also gives access to the images of mean lifetime of the
direct electron-positron annihilations and the probability of the positronium
formation~\cite{patentMoskal}. Establishing correlations of these parameters with the cancer
grade requires systematic long-term study. There are first in-vitro measurements indicating
differences in positronium properties in healthy and cancerous tissues~\cite{JasinskaActaB:2017}.
The application of the method advocated in this article makes an in-vivo investigations possible
which, due to the fact that positronium interacts in the living organisms with bio-active molecules,
may reveal yet unknown features useful for diagnosis.  

Finally it is worth noting that one of the important feature of the mean ortho-positronium lifetime
image is that it does not require attenuation corrections~\cite{patentMoskal}. This is because of the lifetime of ortho-positronium is independent of photons attenuation,
and though suppression of photons in the body affects the statistics in a given voxel, it does not affect the shape of the lifetime spectrum, leaving the mean lifetime unaltered.

\begin{backmatter}
\section*{Ethics approval and consent to participate}
Not applicable
\section*{Consent for publication}
Not applicable
\section*{Availability of data and material}
The data that support the findings of this study are available from the corresponding author upon
 request.
\section*{Competing interests}
The authors declare that they have no competing interests
\section*{Funding}
This work was supported by The Polish National Center for Research
and Development through grant INNOTECH-K1/IN1/64/159174/NCBR/12, the
Foundation for Polish Science through the MPD and TEAM POIR.04.04.00-00-4204/17 programmes, the National Science Centre of Poland through grants no.\
2016/21/B/ST2/01222, 2017/25/N/NZ1/00861, the Ministry for Science and Higher Education through grants no. 6673/IA/SP/2016,
7150/E-338/SPUB/2017/1 and 7150/E-338/M/2017, 7150/E-338/M/2018 and the Austrian Science Fund FWF-P26783.
\section*{Authors' contributions}
All authors contributed to the manuscript discussion, interpretation of results, correction and approval and elaboration of the simulation, analysis and reconstruction methods used to achieve the described result.
PM has conceived the method, derived the analytic estimation of sensitivity, and was a major contributor in writing the manuscript. 
DK prepared the Monte Carlo 
simulations and analyzed positronium mean lifetime distributions.
RS provided reconstruction with TOF-FBP algorithm.

\section*{Acknowledgements}
The authors acknowledge technical and administrative support of A.~Heczko, M.~Kajetanowicz and W.~Migda\l{}.

\bibliographystyle{vancouver}
\bibliography{jpet_imaging}
\end{backmatter}
\end{document}